\documentclass[12pt]{article}
\usepackage{epsfig}
\usepackage{graphicx}
\usepackage{amsmath}

\begin{document}

 \newcommand{\beq}{\begin{equation}}
\newcommand{\eeq}{\end{equation}}
\newcommand{\bea}{\begin{eqnarray}}
\newcommand{\eea}{\end{eqnarray}}
\newcommand{\beqn}{\begin{eqnarray}}
\newcommand{\eeqn}{\end{eqnarray}}
\newcommand{\beas}{\begin{eqnarray*}}
\newcommand{\eeas}{\end{eqnarray*}}
\newcommand{\defi}{\stackrel{\rm def}{=}}
\newcommand{\non}{\nonumber}
\newcommand{\bquo}{\begin{quote}}
\newcommand{\enqu}{\end{quote}}
\newcommand{\qt}{\tilde q}
\newcommand{\m}{\tilde m}
\newcommand{\trho}{\tilde{\rho}}
\newcommand{\tn}{\tilde{n}}
\newcommand{\tN}{\tilde N}
\newcommand{\gsim}{\lower.7ex\hbox{$\;\stackrel{\textstyle>}{\sim}\;$}}
\newcommand{\lsim}{\lower.7ex\hbox{$\;\stackrel{\textstyle<}{\sim}\;$}}


\def\de{\partial}
\def\Tr{ \hbox{\rm Tr}}
\def\const{\hbox {\rm const.}}
\def\o{\over}
\def\im{\hbox{\rm Im}}
\def\re{\hbox{\rm Re}}
\def\bra{\langle}\def\ket{\rangle}
\def\Arg{\hbox {\rm Arg}}
\def\Re{\hbox {\rm Re}}
\def\Im{\hbox {\rm Im}}
\def\diag{\hbox{\rm diag}}


\def\QATOPD#1#2#3#4{{#3 \atopwithdelims#1#2 #4}}
\def\stackunder#1#2{\mathrel{\mathop{#2}\limits_{#1}}}
\def\stackreb#1#2{\mathrel{\mathop{#2}\limits_{#1}}}
\def\Tr{{\rm Tr}}
\def\res{{\rm res}}
\def\Bf#1{\mbox{\boldmath $#1$}}
\def\balpha{{\Bf\alpha}}
\def\bbeta{{\Bf\beta}}
\def\bgamma{{\Bf\gamma}}
\def\bnu{{\Bf\nu}}
\def\bmu{{\Bf\mu}}
\def\bphi{{\Bf\phi}}
\def\bPhi{{\Bf\Phi}}
\def\bomega{{\Bf\omega}}
\def\blambda{{\Bf\lambda}}
\def\brho{{\Bf\rho}}
\def\bsigma{{\bfit\sigma}}
\def\bxi{{\Bf\xi}}
\def\bbeta{{\Bf\eta}}
\def\d{\partial}
\def\der#1#2{\frac{\d{#1}}{\d{#2}}}
\def\Im{{\rm Im}}
\def\Re{{\rm Re}}
\def\rank{{\rm rank}}
\def\diag{{\rm diag}}
\def\2{{1\over 2}}
\def\ntwo{${\mathcal N}=2\;$}
\def\nfour{${\mathcal N}=4\;$}
\def\none{${\mathcal N}=1\;$}
\def\ntwot{${\mathcal N}=(2,2)\;$}
\def\ntwoo{${\mathcal N}=(0,2)\;$}
\def\x{\stackrel{\otimes}{,}}

\def\ba{\beq\new\begin{array}{c}}
\def\ea{\end{array}\eeq}
\def\be{\ba}
\def\ee{\ea}
\def\stackreb#1#2{\mathrel{\mathop{#2}\limits_{#1}}}

\def\Tr{{\rm Tr}}
\newcommand{\cpn}{CP$(N-1)\;$}
\newcommand{\wcpn}{wCP$_{N,\tilde{N}}(N_f-1)\;$}
\newcommand{\wcpd}{wCP$_{\tilde{N},N}(N_f-1)\;$}
\newcommand{\vp}{\varphi}
\newcommand{\pt}{\partial}
\newcommand{\ve}{\varepsilon}
\renewcommand{\theequation}{\thesection.\arabic{equation}}

\setcounter{footnote}0

\vfill

\begin{titlepage}

\begin{flushright}
FTPI-MINN-12/13, UMN-TH-3040/12\\
\end{flushright}

\vspace{2mm}

\begin{center}
{  \Large \bf  
Confronting Seiberg's  Duality with \boldmath{$r$} Duality  
\\[2mm] 
in  \boldmath{\none} 
Supersymmetric QCD
}

\vspace{5mm}

 {\large \bf    M.~Shifman$^{\,a}$ and \bf A.~Yung$^{\,\,a,b}$}
\end {center}

\begin{center}

$^a${\it  William I. Fine Theoretical Physics Institute,
University of Minnesota,
Minneapolis, MN 55455, USA}\\[1mm]
$^{b}${\it Petersburg Nuclear Physics Institute, Gatchina, St. Petersburg
188300, Russia
}
\end{center}

\vspace{3mm}

\begin{center}
{\large\bf Abstract}
\end{center}

Systematizing our results on $r$ duality obtained previously we focus on comparing
$r$ duality with the generalized Seiberg duality in the $r$ vacua of \ntwo and \none
super-Yang--Mills theories with the
U($N$) gauge group and $N_f$ matter flavors  ($N_f>N$). The number of condensed (s)quarks $r$ is assumed to be
in the interval $\frac{2}{3}N_f < r \le N$. To pass to \none 
we introduce an  \ntwo-breaking deformation, 
a  mass term $\mu$ for the adjoint matter, eventually decoupling the adjoint matter in the limit of large $\mu$.
If one starts from a
 large value of the parameter $\xi\sim\mu m$, where the original theory
 is at weak coupling, and let $\xi$ decrease  one hits a 
 a crossover transition from weak to strong coupling (here $m$ is a typical value of the quark masses). 
Below this transition the original theory
 is described in terms of a weakly coupled infrared-free $r$ dual
 theory with the  U$(N_f-r)$ gauge group and  $N_f$ light quark-like dyon flavors.
 Dyon condensation leads to confinement 
of monopoles, defying a naive expectation of  quark confinement. 
The quarks and gauge bosons of the original theory are in  an
``instead-of-confinement'' phase. The $r$ and Seiberg dualities are demonstrated to coincide 
in the  $r=N$ vacua.  In the  $\frac{2}{3}N_f <r<N$ vacua two dualities do not match. In this window
Seiberg's dual is at strong coupling while our $r$-dual model is at weak coupling. 
Thus, we can speak of triality. 
Seiberg's dual solution  at weak coupling reappears again at $r<N_f-N<\frac 13 N_f$.

\vspace{2cm}

\end{titlepage}

 \newpage



\section {Introduction }
\label{intro}
\setcounter{equation}{0}

The discovery of the Seiberg duality \cite{Sdual,IS,msobzor} was a major breakthrough in \none 
Yang--Mills theories at strong coupling, with far reaching consequences both in field theory and string theory. 
In this paper we will explore interrelations between the
Seiberg duality and a novel, recently discovered $r$ duality, in those situations where they overlap.

The original Seiberg duality in Yang--Mills theories with matter was most useful inside the conformal window, in the
conformal regime. Our prime interest is in theories with confinement. The initial impetus for explorations of confinement in 
supersymmetric Yang--Mills was given by the 
Seiberg-Witten solution \cite{SW1,SW2} revealing condensation 
of monopoles \cite{mandelstam} in the monopole vacua of \ntwo supersymmetric QCD. The mechanism of 
string formation and confinement obtained in  \cite{SW1,SW2}
 is essentially Abelian \cite{DS,HSZ,Strassler,VY}. 
 
The non-Abelian gauge group (say, SU(2)) is broken down to an Abelian subgroup  at a high scale 
by condensation of 
the adjoint scalars. An effective Abelian low-energy theory ensues.
The monopole condensation and formation of confining flux tubes (strings) occurs in
this effective  Abelian theory. 

Within Seiberg--Witten solution it remained unclear in which way a confining
 scenario could work in \none QCD, where there are no adjoint scalars and 
no dynamical Abelization. Attempts to extrapolate the line of reasoning
of  \cite{SW1,SW2} to \none QCD   were hindered due to the fact
that the low-energy theory in the monopole vacua becomes strongly 
coupled and untreatable by known methods.

In a bid to uncover a non-Abelian implementation of confinement we passed 
to the quark vacua of \ntwo supersymmetric QCD with the U$(N)$ gauge group and $N_f$ flavors ($N_f>N$).
In this setting not only non-Abelian strings were constructed \cite{HT1,ABEKY,SYmon} but, as an additional bonus, 
 continuation to \none SQCD became possible \cite{SYdual,SYN1dual,SYrvacua}.
 To this end we deformed \ntwo SQCD by adding a mass term $\mu$ for the adjoint matter. On the way from small to 
 large $\mu$ an ``instead-of-confinement'' phase sets in. 
 We found a crossover  (in the Fayet--Iliopoulos \cite{FI} parameter $\xi$ ), a transition that takes us
 from weak  to strong coupling in \none SQCD,
 and established a dual (weakly coupled) theory in the regime where the original \none SQCD is strongly coupled.
 Thus, we observed what can be called $r$ duality in \none.
 
 To be more exact, in our previous paper \cite{SYrvacua}  
 in which all necessary technical work was carried out, 
 we explored the $r$ vacua of the theory, with\,\footnote{Our definition of $r$ refers to the 
 domain of large quark masses, see Sec.~\ref{outl}. It will become clear in Sec.~\ref{GK} that
 effectively $r$ depends on the quark masses.
 }
\beq
\frac{2}{3}N_f < r \le N\,. 
\label{rrange}
\eeq
This explains the origin of the term, $r$ duality. At the same time, the original
 Seiberg duality (formulated in \cite{Sdual,IS} in the  monopole $r=0$ vacua) can be generalized to $r$ vacua  \cite{CKM}
which survive in passing from \ntwo to our basic \none  model at large but finite $\mu$.\footnote{  
At $\mu = \infty$ only the $r=0$ monopole vacua remain, while others become run-away vacua.}
Further explorations of the generalized Seiberg duality were undertaken in \cite{GivKut}. In these  works 
classical vacua were identified -- the vacua that correspond to Seiberg dual description.

Thus, our $r$ duality in the $r$ vacua can (and should) be compared with Seiberg's duality. Are they identical or complementary? How can they coexist?

These are the questions we address here building on the technical work carried out previously.  We will prove that
at $r=N$  both dualities present one and the same description. This is not always the case, however. In the window
 $\frac{2}{3}N_f <r<N$ Seiberg's dual is a model at strong coupling and, thus, is of a limited  use  from the standpoint of
 description of low-energy physics.  At the same time, our $r$-dual model is at 
 weak coupling (in fact, infrared free), and thus fully describes 
 low-energy physics. In this window we can speak of detection of a triality, conceptually similar to
 that found in \cite{KINS} in SO$(N)$ model: two of the dual models in a 
 triplet are strongly coupled while the third one is weakly coupled. 
 
 We will argue that among the Seiberg dual solutions found in 
 \cite{GivKut} the ones that are at weak coupling
  refer to the following domain of $r$:
\beqn
r&<&\tilde{N}\,,
\nonumber\\[1mm]
\tilde{N} &\equiv& N_f - N\,.
\label{12}
\eeqn

\begin{figure}[h]
\epsfxsize=8cm
\centerline{\epsfbox{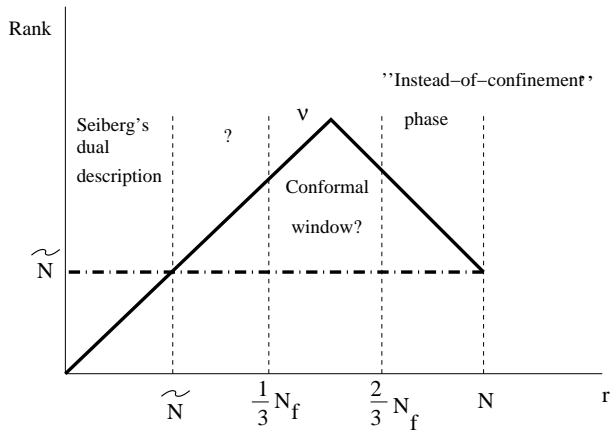}}
\caption{\small Rank of the dual gauge groups in the triality triplet  is
plotted as a function of $r$.  For $r$ duality this rank $=\nu$ (solid line, see Eq.~(\ref{nu})) while
 for the Seiberg duality the rank of the dual gauge group is $\tN$ for all $r$, dashed-dotted line. The domains
of the weakly coupled Seiberg's dual  (the leftmost strip) and so far established  $r$ duality,
i.e. ``instead-of-confinement'' phase (the rightmost strip) do not overlap except the fact that the 
identical coincidence 
between Seiberg's dual and ours occurs
at  $r=N$. }
\label{figdual}
\end{figure}

To explain  
the interrelation between our $r$ duality and  Seiberg's duality it is instructive to look at Fig.~\ref{figdual}.
Our derivation \cite{SYrvacua} based on exact results \cite{SW1,SW2} for \ntwo$\!\!$, with the subsequent (theoretically controlled)
continuation to \none$\!\!$, refers to the rightmost strip. 
Our solution is fully controllable, and the dual model we get is at weak coupling,
while Seiberg's duality in this domain is not at weak coupling (except its very boundary, $r=N$).
Weak coupling regime in the Seiberg's  dual theory refers to the leftmost strip.
What lies between these two strips?

Strictly speaking, today we do not know for certain. One can present certain  
speculations reflected in the central area in Fig.~\ref{figdual}, see also \cite{SYrvacua}.
These speculations go beyond the scope of the present paper. One can consider this question as a task for a future investigation. 

Another problem for a future analysis is interpreting $r$ duality in the framework of strings/branes, in the spirit
it had been done with the Seiberg duality.

The paper is organized as follows. 
Section~\ref{outl} highlights the main points of our analysis.
In Sec.~\ref{bulk} we briefly describe the basic theory we work with, $\mu$-deformed \ntwo SQCD. 
In Sec.~\ref{rdual} we review $r$ duality and  ``instead-of-confinement'' mechanism in the $r$ vacua. In 
Sec.~\ref{rduallargemu} we review $r$ duality at large $\mu$ in the \none SQCD limit. In Sec.~\ref{seiberg} 
we describe the
generalized Seiberg duality and compare it with
  our $r$ duality in the $r=N$ vacuum. 
  Section~\ref{GK} presents an analysis of the results obtained in \cite{GivKut}.
  In this section we establish that the Seiberg dual solutions found in \cite{GivKut} are at weak coupling only 
in the domain (\ref{12}), i.e. at $r< \tN$. 
   In Sec.~\ref{CSW} we confront our $r$ duality with that of Seiberg in the
$r<N$ vacua. We argue that Seiberg's duality   is not implemented at weak coupling
at $\frac 23 N_f < r <N$, while the $r$ duality is. Finally, in Appendix 
 a more general $\mu$ deformation is considered.

\section{Analysis outline and main statements}
\label{outl}
\setcounter{equation}{0}

As was, mentioned, our starting point is \ntwo SQCD, with the U$(N)$ gauge group,
in which  we choose vacua in a judicious way. First we
treat it at
  large values of an effective Fayet--Iliopoulos (FI) parameter $\xi$, namely,  $\xi\sim \mu m$, where $m$ is a 
  generic quark mass.     At large $m$ we arrange 
$r$ quark flavors to condense. This is our definition of the parameter $r$. In fact, the number of condensed quarks
can depend on $m$, see Sec. \ref{GK} for details.

In the large-$m$ vacuum with $r$ condensed quarks
the effective low energy-theory
with the gauge group U$(r)\times$U(1)$^{N-r}$ is at weak coupling.  

A global color-flavor locked symmetry survives in the limit of the equal quark masses.
At large $\xi$ this theory 
supports non-Abelian flux tubes (strings)
 \cite{HT1,ABEKY,SYmon,HT2} (see also \cite{Trev,Jrev,SYrev,Trev2} for reviews). It is the
formation of these strings that ensures confinement of monopoles. Monopoles manifest themselves 
 as two-string junctions.  
The distinction between the  $r<N$ vacua and that  with $r=N$ is that for   $r<N $ one U(1) factor 
of the U$(N)$ gauge group always remains unbroken \cite{Cachazo2}. Thus, in this case, 
long-range forces are always present.

Exploring these vacua we established  an  $r$-duality. 
Upon reducing the  $\xi$  parameter the theory under consideration
goes through a crossover transition \cite{SYdual,SYN1dual,SYrvacua} into a strongly 
coupled regime which can be described
in terms of a {\em weakly coupled  dual} infrared-free \ntwo SQCD.
 The gauge group of the dual theory is 
\beq
U(\nu)\times U(1)^{N-\nu}, \qquad
\nu=\left\{
\begin{array}{cc}
r, & r\le \frac{N_f}{2}\\[2mm]
N_f-r, & r > \frac{N_f}{2}, \\
\end{array}
\right.
\label{nu}
\eeq
So far we limited ourselves to the case $\nu=N_f-r$.  For $r=N$ vacuum our  $r$ dual gauge group reduces to 
that of Seiberg's duality, in which the first factor in the second line of (\ref{nu}) is U$(N_f-N)$.
 The coincidence does not extend to the case  $r<N$  \cite{SYrvacua}: instead of U$(N_f-N)$ we get U$(N_f-r)$.

\vspace{1mm}

It is worth noting  that the presence of the 
non-Abelian SU$(\nu)\times $U(1)$^{N_f-\nu}$ gauge group at the roots of 
the nonbaryonic branches in massless ($\xi=0$)  \ntwo SU$(N)$ SQCD was first observed in \cite{APS}. 
Moreover, 
in this paper the SU$(N_f-N)$ dual  gauge group  was identified
at the root of the baryonic Higgs branch in the  SU($N$) theory.
The relation between
$r$ and $\nu$ given by (\ref{nu}) was  noted in \cite{Ookouchi,BolKonMar}, where it was interpreted as a correspondence
between the ``classical and quantum $r$ vacua.'' We interpret it as a duality occurring upon reducing $\xi$ 
 below the crossover transition line.

The dual theory supports non-Abelian strings due to condensation of light dyons in  much the same way
as non-Abelian strings in the original theory which are due to condensation of quarks. 
The strings  of the dual theory confine {\em monopoles} too, 
rather than quarks \cite{SYdual,SYrvacua}.
This is explained by the fact that the  light dyons   condensing in the dual theory  
carry weight-like chromoelectric charges (in addition to chromomagnetic charges). In other words, they carry
the quark charges. The strings formed through condensation of these dyons
can confine only  the states with the root-like magnetic charges, i.e. 
the monopoles  \cite{SYdual}.  

Thus, 
our $r$ duality is {\em not}  electromagnetic. 
 There  is no confinement of the chromoelectric charges in the dual theory; on the contrary, they
are Higgs-screened.

\begin{figure}
\epsfxsize=6cm
\centerline{\epsfbox{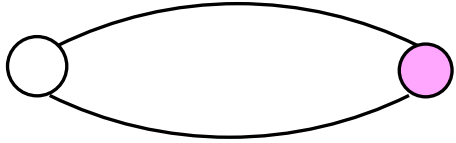}}
\caption{\small Mesons built from the monopole-antimonopole pairs connected by two strings.
Open and closed circles denote the monopole and antimonopole, respectively.}
\label{figmeson}
\end{figure}

At strong coupling, when the $r$  dual description sets in, the gauge bosons
and quarks of the original theory   are in what we call  ``instead-of-confinement'' phase. Namely,
the quarks and
gauge bosons of the original theory  decay into monopole-antimonopole 
pairs on the curves of marginal stability (CMS) \cite{SYdual,SYtorkink}.
The (anti)monopoles forming the pair are confined. In other words, the original quarks and gauge bosons 
evolve in the 
strong coupling domain of small $\xi$  into stringy mesons with two constituents 
being connected by two strings as shown in  Fig.~\ref{figmeson}.  
These mesons are expected to lie on the Regge trajectories.

Moreover, deep in the non-Abelian quantum regime the confined mono\-poles were demonstrated \cite{SYtorkink} 
to belong to the {\em fundamental representation} of the global  (color-flavor locked) group. 
Therefore, the  monopole-antimonopole
mesons can be both, in the adjoint and singlet representation of this  group. 
This  pattern  seems to be similar to what we have in the real world. 
The role of the ``constituent quarks'' inside the above mesons is played by the monopoles.

At this stage we are still not far away from the \ntwo limit. Then
 we increased the deformation parameter $\mu$  decoupling the
adjoint fields thus sending the original theory  to the limit of \none SQCD \cite{SYN1dual,SYrvacua}. 
In the passage from \ntwo to \none   we observed no dramatic qualitative changes.
At large $\mu$ the dual theory was demonstrated to be   weakly coupled and infrared  free, with the 
 U$(\nu)$  gauge group  and $N_f$ light dyons $D^{lA}$, (here $l=1, ... ,\nu$ is the color index 
in the dual gauge group, while $A=1,...,N_f$ is the flavor index). 
Non-Abelian strings still confine monopoles. ``Instead-of-confinement'' mechanism works at large $\mu$ as follows.
In the $r=N$ vacuum the quarks and gauge bosons of  the original \none SQCD  continue to be
presented by stringy mesons built from the
monopole-antimonopoles pairs connected by two non-Abelian strings, see Fig.~\ref{figmeson}.

\vspace{2mm}

In the  $r<N$ vacua (but $r> \frac 23 N_f$) there is a novel feature: one (say, $N$-th) $Z_N$ string is absent
in $r<N$ vacua and the associated flux of the unbroken U(1)$^{\rm unbr}$ gauge factor is not squeezed into a flux tube. 
It is spread out in space via the Coulomb law.

\begin{figure}
\epsfxsize=8cm
\centerline{\epsfbox{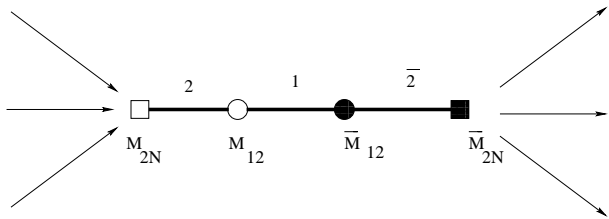}}
\caption{\small Example of the dipole meson formed as result of breaking of the second string by pair creation of 
the monopole
$M_{2N}$ (shown by boxes) interpolating between the second   string and the would-be $N$-th string, which is absent.
Arrows denote unconfined flux.
Circles denote the monopoles $M_{KK'}$, $K,K'=1, ... ,\nu$. 
Open and closed circles/boxes denote  the monopoles and antimonopoles, respectively.}
\label{figdipole}
\end{figure}

As a result,  non-Abelian strings become metastable in the $r<N$ vacua: they can be broken by a monopole-antimonopole
pair creation. The monopoles in the produced pair are junctions of  one of the first $r$ $Z_N$-strings 
with  the would-be 
$N$-th string (which is in fact absent). An example of the meson resulting in this way is shown in Fig.~\ref{figdipole}. 
The endpoints emit fluxes of the unbroken U(1) gauge field. This makes this meson a dipole-like configuration. 

\vspace{2mm} 
Note, that the non-Abelian fluxes of the SU$(\nu)$ gauge group are always squeezed in the non-Abelian strings.
Long-range forces are associated only with the unbroken U(1)$^{\rm unbr}$ gauge factor.
The monopoles inside  the dipole meson cannot annihilate
if the overall flavor charge of the meson is nontrivial, say, the meson is in the adjoint.

Armed with the knowledge of the confining dynamics in the dual pair of \none theories,
we move on to compare our $r$ duality with Seiberg's duality. The simplest case is $r=N$.
In the $r=N$ vacuum our dual gauge group U$(\nu=N_f-r)$ coincides with Seiberg's dual group U$(\tN)$, where
\beq
\tN=N_f-N.
\label{tN}
\eeq
 Moreover, in this case the generalized Seiberg dual superpotential has a classical vacuum. 
 We show that,
upon integrating out heavy mesonic $M$-fields,  this superpotential coincides with our $r$-dual superpotential 
obtained in \cite{SYN1dual}.  Seiberg's ``dual quarks'' are found to reduce to  quark-like dyons $D^{lA}$, 
up to a normalization. Both dualities perfectly match in the  $r=N$ vacuum. This entails, in particular,
 that in the $r=N$ vacuum Seiberg's
``dual quarks'' are quark-like dyons, rather than  monopole-like states. Their condensation 
leads to confinement of monopoles, while the quarks are in the ``instead-of-confinement'' 
phase \cite{SYN1dual}.

For $\frac{2}{3}N_f <r<N$ the generalized Seiberg superpotential has no supersymmetric 
classical vacua provided that the  quark masses are generic. 
However, there are so-called ``quantum vacua'' which can be found by integrating
out Seiberg's ``dual quarks'' \cite{IS,IS2} which, in turn,  leads to an  effective superpotential 
in terms  of mesonic $M$ fields.

In doing so one obtains an extension of the Afleck--Dine--Seiberg  (ADS) superpotential 
\cite{ADS} to $N_f>N$.
The latter correctly reproduces the quark and gaugino condensates. We explicitly  check that 
it gives the same results for 
the chiral condensates as the exact analysis of the chiral rings carried out in \cite{Cachazo2}.

At the same time, in the  $\frac{2}{3}N_f <r<N$ vacua our $r$ duality does not match Seiberg's duality. 
We demonstrate our dual theory to have the
U$(\nu)$ gauge group instead of U$(\tN)$ and a
different superpotential for light matter. 
Our dual theory does have a supersymmetric classical vacuum
 and, in a certain regime (with small $\xi$), stays at weak coupling.
Our interpretation of this is as follows. In the range $\frac{2}{3}N_f < r < N$ 
 generalized Seiberg dual theory does 
{\em not} describe low-energy physics in its entirety in the $r$ vacua. 
However, it does describe the chiral sector in the sense of the
Veneziano--Yankielowicz effective superpotential \cite{Ven} (which is not a genuine
low-energy superpotential). The spectrum of excitations is not  reproduced correctly.

Low-energy physics in the  $r$ vacua is described (in the range $\frac{2}{3}N_f < r < N$)  
by $r$ duality, with the dual gauge group U$(\nu=N_f-r)$ replacing Seiberg's U$(\tN=N_f-N)$.

We also show that for smaller $r$, namely for $r<\tN$, Seiberg's dual theory has
 supersymmetric classical vacua
and in fact describes   low-energy physics. This range, however, is beyond the scope of the present paper. 

\section {\boldmath{$\mu$}-Deformed \boldmath{\ntwo} SQCD and its vacuum structure at large 
\boldmath{$\xi$}}
\label{bulk}
\setcounter{equation}{0}

The  model we start from has the   
U($N$)=SU$(N)\times$U(1) gauge symmetry  and $N_f$ massive quark hypermultiplets. In the absence
of the $\mu$  deformation the model  is \ntwo  supersymmetric. We assume that
$N_f>N$ but $N_f<\frac32 N$.  The latter inequality ensures infrared freedom of the dual theory. 

In addition, we will introduce the mass term $\mu$ 
for the adjoint matter breaking \ntwo supersymmetry down to \none. 

The field content is as follows. The \ntwo vector multiplet
consists of the  U(1)
gauge field $A_{\mu}$ and the SU$(N)$  gauge field $A^a_{\mu}$,
where $a=1,..., N^2-1$, and their Weyl fermion superpartners plus
complex scalar fields $a$, and $a^a$ and their Weyl superpartners, respectively.
The $N_f$ quark multiplets of  the U$(N)$ theory consist
of   the complex scalar fields
$q^{kA}$ and $\tilde{q}_{Ak}$ (squarks) and
their   fermion superpartners --- all in the fundamental representation of 
the SU$(N)$ gauge group.
Here $k=1,..., N$ is the color index
while $A$ is the flavor index, $A=1,..., N_f$. We will treat $q^{kA}$ and $\tilde{q}_{Ak}$
as rectangular matrices with $N$ rows and $N_f$ columns. 

Let us first discuss the undeformed  \ntwo theory.
 The  superpotential has the form
 \beq
{\mathcal W}_{{\mathcal N}=2} = \sqrt{2}\,\sum_{A=1}^{N_f}
\left( \frac{1}{ 2}\,\tilde q_A {\mathcal A}
q^A +  \tilde q_A {\mathcal A}^a\,T^a  q^A + m_A\,\tilde q_A q^A\right)\,,
\label{superpot}
\eeq
where ${\mathcal A}$ and ${\mathcal A}^a$ are  chiral superfields, the ${\mathcal N}=2$
superpartners of the gauge bosons of  U(1) and SU($N$), respectively, while $m_A$ are the quark masses.
Next, we add the mass term for the adjoint fields which breaks \ntwo
supersymmetry down to \none,
\beq
{\mathcal W}_{{\rm br}}=\sqrt{\frac{N}{2}}\,\frac{\mu_0}{2} {\mathcal A}^2
+  \frac{\mu}{2}({\mathcal A}^a)^2,
\label{msuperpotbrg}
\eeq
where $\mu_0$ and $\mu$ is are mass parameters for the chiral
superfields in \ntwo gauge supermultiplets,
U(1) and SU($N$), respectively. In the bulk o the paper we will consider the single trace perturbation which amounts to choosing $\mu_0$ in such a way that the parameter
\beq
\gamma = 1-\sqrt{\frac2N}\frac{\mu_0}{\mu}
\label{gamma}
\eeq
vanishes.
In this case the deformation superpotential (\ref{msuperpotbrg}) reduces to a single trace,
\beq
{\mathcal W}_{{\rm br}}=
  \mu\,{\rm Tr}\,\Phi^2,
\label{msuperpotbr}
\eeq
where
\beq
\Phi=\frac12\, {\mathcal A} + T^a\, {\mathcal A}^a.
\label{Phi}
\eeq
Non-single trace deformation is discussed in the Appendix.

The mass term (\ref{msuperpotbr}) splits the 
\ntwo supermultiplets, breaking
\ntwo supersymmetry down to \none. 
Our strategy is as follows. First we assume that deformation to be weak,
\beq
|\mu | \ll \Lambda_{{\mathcal N}=2}\, ,
\label{smallmu}
\eeq
where $\Lambda_{{\mathcal N}=2}$ is the scale of the \ntwo  theory,
so the theory is close to the \ntwo limit. We reduce the parameter $\xi$
and describe $r$ duality at small $\xi$
\cite{SYdual,SYrvacua}. Finally, we make $\mu$ large sending the theory to \none SQCD,
 and discuss how this affects the dual theory \cite{SYN1dual,SYrvacua}. 
 
\subsection{The \boldmath{$r=N$} vacuum}

With generic values of the quark masses we have 
\beq
{\cal N}_{r=N} = C_{N_f}^{N}= \frac{N_f!}{N!(N_f-N)!}
\label{numva}
\eeq
 isolated vacua in which $r=N$ quarks (out of $N_f$) develop
vacuum expectation values  (VEVs).
Following \cite{SYdual} consider, say, the vacuum in which the first $N$ flavors develop VEVs, to be denoted as (1, 2 ..., $N$).
In this vacuum  the
adjoint fields  develop  
VEVs too, namely,
\beq
\left\langle \Phi\right\rangle = - \frac1{\sqrt{2}}
 \left(
\begin{array}{ccc}
m_1 & \ldots & 0 \\
\ldots & \ldots & \ldots\\
0 & \ldots & m_N\\
\end{array}
\right).
\label{avev}
\eeq

For generic values of the quark masses, the  SU$(N)$ subgroup of the gauge 
group is
broken down to U(1)$^{N-1}$. However, in the {\em special limit}
\beq
m_1=m_2=...=m_{N_f},
\label{equalmasses}
\eeq
the  adjoint field VEVs do not break the SU$(N)\times$U(1) gauge group.
In this limit the theory acquires  a global flavor SU$(N_f)$ symmetry.

With all quark masses equal (and to the leading order in $\mu$)
 the mass term for the adjoint matter (\ref{msuperpotbr})
reduces to the Fayet--Iliopoulos $F$-term of the U(1) factor of the SU$(N)\times$U(1) gauge group,
  which does {\em not} break \ntwo supersymmetry \cite{HSZ,VY}. 
Higher orders in the parameter $\mu$
  break \ntwo supersymmetry by splitting all \ntwo multiplets.
If the quark masses are unequal the U($N$) gauge group is broken  down to U(1)$^{N}$
by the adjoint field VEVs (\ref{avev}). 

Using (\ref{msuperpotbr}) and (\ref{avev}) it is not difficult to obtain the quark field VEVs
 from Eq.~(\ref{superpot}) supplemented by $D$-term conditions.   By virtue of a gauge rotation they can be written as \cite{SYfstr}
\beqn
\langle q^{kA}\rangle &=& \langle\bar{\tilde{q}}^{kA}\rangle=\frac1{\sqrt{2}}\,
\left(
\begin{array}{cccccc}
\sqrt{\xi_1} & \ldots & 0 & 0 & \ldots & 0\\
\ldots & \ldots & \ldots  & \ldots & \ldots & \ldots\\
0 & \ldots & \sqrt{\xi_N} & 0 & \ldots & 0\\
\end{array}
\right),
\nonumber\\[4mm]
k&=&1,..., N\,,\qquad A=1,...,N_f\, ,
\label{qvev}
\eeqn
where we present the quark fields as  matrices in color ($k$) and flavor ($A$) indices.
The Fayet--Iliopoulos $F$-term parameters for each U(1) gauge factor are given (in the quasiclassical
approximation) by the following expressions:
\beq
\xi_P \approx 2\;\mu m_P,
\qquad P=1,...,N.
\label{xis}
\eeq

\vspace{2mm}

While the adjoint VEVs do not break the SU$(N)\times$U(1) gauge group in the limit
(\ref{equalmasses}), the quark condensate (\ref{qvev}) does result in  the spontaneous
breaking of both gauge and flavor symmetries.
A diagonal global SU$(N)$ combining the gauge SU$(N)$ and an
SU$(N)$ subgroup of the flavor SU$(N_f)$
group survives, however.

Thus, the pattern  of the
color and flavor symmetry breaking
is 
\beq
{\rm U}(N)_{\rm gauge}\times {\rm SU}(N_f)_{\rm flavor}\to  
{\rm SU}(N)_{C+F}\times  {\rm SU}(\tN)_F\times {\rm U}(1)\,,
\label{c+f}
\eeq
where $\tN$ is given by (\ref{tN}).
Here SU$(N)_{C+F}$ is a global unbroken color-flavor rotation, which involves the
first $N$ flavors, while the SU$(\tN)_F$ factor stands for the flavor rotation of the 
last $\tN$ quarks.
The presence of the global SU$(N)_{C+F}$ group is instrumental for
formation of the non-Abelian strings \cite{HT1,ABEKY,SYmon,HT2,SYfstr}.
Tensions of $N$ elementary strings are determined by parasmeters $\xi_P$ via \cite{SYfstr}
\beq
T_P=2\pi\xi_P.
\label{ten}
\eeq
These strings confine monopoles, in fact elementary monopoles become junctions of two distinct elementary strings
\cite{Tong,SYmon,HT2}.

Since the global (flavor) SU$(N_f)$ group is broken by the quark VEVs anyway, it will be helpful for 
our purposes
to consider
the following mass splitting:
\beq
m_P=m_{P'}, \qquad m_K=m_{K'}, \qquad m_P-m_K=\Delta m
\label{masssplit}
\eeq
where 
\beq
 P, P'=1, ... , N\,\,\,\, {\rm and}   \,\,\,\, K, K'=N+1, ... , N_f\,.
 \label{pppp}
\eeq
This mass splitting respects the global
group (\ref{c+f}) in the $(1,2,...,N)$ vacuum. Moreover, this vacuum  becomes  isolated.
No Higgs branch  develops.  We will often assume this limit below.

Now let us briefly discuss the  perturbative excitation spectrum. 
Since both U(1) and SU($N$) gauge groups are broken by the squark condensation, all
gauge bosons become massive. To the leading order in $\mu$, \ntwo supersymmetry 
is unbroken.  In fact, with nonvanishing $\xi_P$'s (see Eq.~(\ref{xis})), both the quarks and adjoint scalars  
combine  with the gauge bosons to form long \ntwo supermultiplets \cite{VY},  for a review see \cite{SYrev}.
In the limit (\ref{masssplit}) $\xi_P\equiv\xi\,,$  and all states come in 
representations of the unbroken global
 group (\ref{c+f}), namely, in the singlet and adjoint representations
of SU$(N)_{C+F}$,
\beq
(1,\, 1), \quad (N^2-1,\, 1)\,,
\label{onep}
\eeq
 and in the bifundamental representations
\beq
 \quad (\bar{N},\, \tN), \quad
(N,\, \bar{\tN})\,.
\label{twop}
\eeq
We mark representations in (\ref{onep}) and (\ref{twop})  with respect to two 
non-Abelian factors in (\ref{c+f}). The singlet and adjoint fields are (i) the gauge bosons, and
(ii) the first $N$ flavors of the squarks $q^{kP}$ ($P=1,...,N$), together with their fermion superpartners.
The bifundamental fields are the quarks $q^{kK}$ with $K=N+1,...,N_f$.
These quarks transform in the two-index representations of the global
group (\ref{c+f}) due to the color-flavor locking. Singlet and adjoint fields have masses of order $g\sqrt{\xi}$,
while masses of bifundamental fields are  $\Delta m$.

The above quasiclassical analysis is valid if the theory is at weak coupling. This is the case if
the quark VEVs are sufficiently large so that the  gauge coupling constant is frozen at a large scale.
From (\ref{qvev}) we see that 
the quark condensates are of order of
$\sqrt{\mu m}$ (see also \cite{SW1,SW2,APS,CKM}). The weak 
coupling condition is
\beq
|\sqrt{\mu m}| \gg\Lambda_{{\mathcal N}=2}\,,
\label{weakcoup}
\eeq
where we assume all quark masses to be of the same order $m_A\sim m$. 
In particular,  the condition (\ref{weakcoup}), combined with the condition (\ref{smallmu}) of smallness
of $\mu$,   implies that the average quark mass $m$ is very large. 

\subsection{The \boldmath{$r<N$} vacua}

At large $\xi$ the quark sector of the theory in the $r$ vacua
 is at weak coupling and can be analyzed semiclassically.
The number of the $r$ vacua with $r<N$  is \cite{CKM}
\beq
{\cal N}_{r<N}=\sum_{r=0}^{N-1} \,(N-r)\,C_{N_f}^{r}= \sum_{r=0}^{N-1}\, (N-r)\,\frac{N_f!}{r!(N_f-r)!}\,.
\label{nurvac}
\eeq
It is equal to the number of choices one can pick up $r$ quarks which develop VEVs (out of $N_f$ quarks) times the
Witten index (number of vacua) in the classically unbroken SU$(N-r)$ pure gauge theory. 

Below  we will consider a
 particular vacuum in which the first $r$ quarks develop VEVs. 
We denote it as ($1, ... ,\, r$). Quasiclassically, with large mass differences,  the adjoint scalar VEVs   are  
\beq
\left\langle {\rm diag}\left(\frac12\, a + T^a\, a^a\right)\right\rangle \approx - \frac1{\sqrt{2}}
\left[m_1,...,m_r,0,...,0 
\right],
\label{avevr}
\eeq
where the first $r$  diagonal elements are  proportional to the
quark masses, while the last $(N-r)$ entries   classically vanish.
In quantum theory they become of order of $\Lambda_{{\mathcal N}=2}$.
The classically unbroken U$(N-r)$ pure gauge sector gets broken through the Seiberg--Witten mechanism \cite{SW1}
first down to U(1)$^{N-r}$ and then almost completely by condensation of $(N-r-1)$ monopoles. A single
 U(1) factor remains unbroken,
because monopoles are charged only with respect to 
the Cartan generators of the SU$(N-r)$ group.
The presence of the unbroken U(1)$^{\rm unbr}$ symmetry
in all $r<N$ vacua makes them different from the $r=N$ vacuum where there are
no long-range forces. In the terminology of \cite{Cachazo2} these sets of vacua belong to 
two different ``branches.''

Following \cite{SYrvacua} we consider the $r=(N-1)$ vacuum as an example.
The low energy theory in 
the $r=N-1$ vacuum at large $\xi$  has non-Abelian gauge fields
$A_{\mu}^{n}$, $n=1, ... , (r^2-1)$ as well as Abelian ones $A_{\mu}$ and $A_{\mu}^{(N^2-1)}$. The last 
field is associated with the last Cartan generator of SU$(N)$. 
These fields have scalar superpartners $a^{n}$, $a$ and 
$a^{(N^2-1)}$. Light matter consists of the $q^{kA}$ quarks, $k=1, ... , r$. Note, that
all non-Abelian gauge fields from the SU$(N)$/SU$(r)$ sector  are heavy and decouple in the large mass limit due to
 the structure of the adjoint VEVs (see (\ref{avevr})). Also the $q^{NA}$ quarks  are heavy and not 
included in the low-energy theory.

The vacuum structure in the $r=N-1$ vacuum is as follows. The adjoint VEVs have the form
\beq
\left\langle {\rm diag} \left( \Phi \right)\right\rangle \approx - \frac1{\sqrt{2}}
\left[\,m_1,...,m_{N-1}, 0\,
\right],
\label{avevN-1}
\eeq
while the (s)quark VEVs are 
\beqn
\langle q^{kA}\rangle &=& \langle\bar{\tilde{q}}^{kA}\rangle=\frac1{\sqrt{2}}\,
\left(
\begin{array}{cccccc}
\sqrt{\xi_1} & \ldots & 0 & 0 & \ldots & 0\\
\ldots & \ldots & \ldots  & \ldots & \ldots & \ldots\\
0 & \ldots & \sqrt{\xi_{N-1}} & 0 & \ldots & 0\\
\end{array}
\right),
\nonumber\\[4mm]
k&=&1,..., (N-1)\,,\qquad A=1,...,N_f\, .
\label{qvevr}
\eeqn
The first  $(N-1)$ parameters $\xi$ are given quasiclassically by (\ref{xis}) while 
\beq
\xi_N =0\,.
\label{xiN}
\eeq
In the $r=N$ vacuum the last entry in (\ref{xis}) is $m_N$  while now we have zero.
The  condition (\ref{xiN}) reflects the fact that the $N$-th quark is heavy and  develops no VEV.
This is also valid in quantum theory \cite{SYrvacua}. 

Quarks interact with a particular linear combination
of the U(1) gauge fields $A_{\mu}$ and $A_{\mu}^{N^2-1}$, namely,
\beq
 A_{\mu} +\sqrt{\frac2{N(N-1)}}\; A_{\mu}^{N^2-1}.
\label{A0}
\eeq
Quark condensation makes this combination massive. The orthogonal combination
\beq
\sqrt{\frac2{N(N-1)}}\; A_{\mu} - A_{\mu}^{N^2-1}.
\label{unbroken}
\eeq
remains massless and corresponds to the unbroken U(1)$^{\rm unbr}$ gauge group.

In the equal mass limit the global flavor symmetry SU$(N_f)$ in the $r$ vacuum is broken 
down to
\beq
{\rm SU}(r)_{C+F}\times  {\rm SU}(\nu=N_f-r)_F\times {\rm U}(1)\,.
\label{c+fr}
\eeq
Now SU$(r)_{C+F}$ is a global unbroken color-flavor rotation, which involves only the
first $r$ flavors, while the SU$(\nu=N_f-r)_F$ factor stands for the flavor rotation of the 
remainder of the  quark set.

Since the global (flavor) SU$(N_f)$ group is broken by the quark VEVs anyway, it is useful  to consider
the  split quark masses, as in (\ref{masssplit}), with (\ref{pppp}) replaced by
\beq
 P, P'=1, ... , r\,\,\,\, {\rm and}   \,\,\,\, K, K'=r+1, ... , N_f\,.
 \label{masssplitr}
\eeq
This mass splitting respects the global
group (\ref{c+fr}) in the $(1,2,...,r)$ vacuum. This vacuum  becomes  isolated.

In much  the same way as in the $r=N$ vacuum in the $r<N$ vacua
   all states in the limit (\ref{masssplitr}) come in 
representations of the unbroken global
 group (\ref{c+fr}), namely, in the singlet and adjoint representations
of SU$(r)_{C+F}$,
\beq
(1,\, 1), \quad (r^2-1,\, 1),
\label{onepr}
\eeq
 and in the bifundamental representations
\beq
 \quad (\bar{r},\, \nu), \quad
(r,\, \bar{\nu})\,.
\label{twopr}
\eeq
 The singlet and adjoint fields are the gauge bosons, and
 the first $r$ flavors of the quarks $q^{kP}$ ($P=1,...,r$). 
The bifundamental fields are the $q^{kK}$ quarks  with $K=r+1, ... , N_f$.
The singlet and adjoint fields have masses of order $g\sqrt{\xi}$, where $\xi$ is the common value of the 
first $r$ $\xi$'s in the limit (\ref{masssplit}), (\ref{masssplitr}). 
The masses of bifundamental fields are  $\Delta m$.

Quasiclassical analysis is valid if the theory is at weak coupling. The weak 
coupling condition in the asymptotically free SU($r$) sector reduces to
\beq
|\sqrt{\xi} | \sim |\sqrt{\mu m}| \gg\Lambda_{{\mathcal N}=2}^{\rm LE}\,,
\label{weakcoupr}
\eeq
where $\Lambda_{{\mathcal N}=2}^{\rm LE}$ is the scale of the low energy   theory  determined by 
\beq
\Lambda_{{\mathcal N}=2}^{2N-N_f}=m^2 \,(\Lambda_{{\mathcal N}=2}^{\rm LE})^{2(N-1)-N_f}\, .
\label{LambdaN2LE}
\eeq

Quarks  in $r=N-1$ vacuum develop VEVs; therefore monopoles should be confined, in much  the same way
as in the $r=N$ vacuum. The distinction is that one U(1) factor of the gauge group
remains unbroken, therefore the associated magnetic flux is unconfined. 
In fact one of $Z_N$ strings (say, the $N$-th string) is absent due to the condition (\ref{xiN}).

Therefore $r$ strings associated with windings
of $r$ quarks can terminate on the monopoles $M_{PN}$, $P=1, ... , r$ interpolating between one of these string and the spurious $N$-th string. The endpoint is a magnetic source of unbroken 
U(1)$^{\rm unbr}$ gauge field. All other monopole fluxes, in particular, all non-Abelian fluxes from the
SU($r$) subgroup are absorbed by confining  strings, see \cite{SYrvacua} for details.

\section{\boldmath{$r$} Duality}
\label{rdual}
\setcounter{equation}{0}

What happens if we   relax the condition (\ref{weakcoup}) or (\ref{weakcoupr}) and pass
to the strong coupling domain at 
\beq
|\sqrt{\xi_P}|\ll \Lambda_{{\mathcal N}=2}\,, \qquad |m_{A}-m_B |\ll \Lambda_{{\mathcal N}=2} 
\label{strcoup}
\eeq
still keeping $\mu$ small? 
 
As was shown in \cite{SYdual,SYrvacua}, our theory  in the $r$ vacuum undergoes a
crossover transition on the way from large to small $\xi$.
The  domain (\ref{strcoup}) 
can be described in terms of weakly coupled (infrared free) dual theory
with with the gauge group
\beq
{\rm U}(\nu)\times {\rm U}(1)^{N-\nu}\,,
\label{rdualgaugegroup}
\eeq
 and $N_f$ light dyon flavors ($r$ is assumed to be be in the range $\frac12 N_f <r\le N$).

\vspace{2mm}

The quark-like dyons $D^{lA}$, 
($l=1, ... ,\nu$, $A=1, ... , N_f$) are in 
the fundamental representation of the SU$(\nu)$ gauge group
 and are charged under the Abelian factors indicated in Eq.~(\ref{rdualgaugegroup}).
 In addition, there are  $(N-\tN)$ or $(r-\nu)$
light quark-like dyons $D^J$, neutral under 
the SU$(\nu)$ group, but charged under the
U(1) factors in the $r=N$ and $r<N$ vacua, respectively. In the $r<N-1$ vacua there are also
$(N-r-1)$ light monopoles charged under the U(1) factors.\footnote{
 We collectively refer to all dyons carrying root-like electric charges    as 
``monopoles.'' This is to avoid confusion with the dyons which appear in Eq.~(\ref{Dvev}). The
latter dyons carry weight-like electric charges and, roughly speaking, behave as
quarks, see \cite{SYdual,SYrvacua} for further details.}

The dyon condensates are  
\beqn
\langle D^{lA}\rangle \! \! &=& \langle \bar{\tilde{D}}^{lA}\rangle =
\!\!
\frac1{\sqrt{2}}\,\left(
\begin{array}{cccccc}
0 & \ldots & 0 & \sqrt{\xi_{1}} & \ldots & 0\\
\ldots & \ldots & \ldots  & \ldots & \ldots & \ldots\\
0 & \ldots & 0 & 0 & \ldots & \sqrt{\xi_{\nu}}\\
\end{array}
\right),
\nonumber\\[4mm]
\langle D^{J}\rangle &=& \langle\bar{\tilde{D}}^{J}\rangle=\sqrt{\frac{\xi_J}{2}}, 
\label{Dvev}
\eeqn
where $J=(\tN+1), ..., N$ in the $r=N$ vacuum and $J=(\nu+1), ..., r$ in the $r<N$ vacua.

The most important feature apparent in (\ref{Dvev}), as compared to the squark VEVs  of the 
original theory (\ref{qvev}),  is a ``vacuum leap'' \cite{SYdual},
\beq
(1, ... ,\, r)_{\sqrt{\xi}\gg \Lambda_{{\mathcal N}=2}} \to (r+1, ... , \,N_f,\,\,(\nu+1), ... ,\, r)_{\sqrt{\xi}\ll \Lambda_{{\mathcal N}=2}}\,.
\label{jump}
\eeq
In other words, if we pick up the vacuum with nonvanishing VEVs of the  first $r$ quark flavors
in the original theory at large $\xi$  and then reduce $\xi$ below 
$\Lambda_{{\mathcal N}=2}$, 
the system goes through a crossover transition and ends up in the vacuum of the dual theory with
the nonvanishing VEVs of the last $\nu $  dyons (plus VEVs of  the SU$ (\nu)$ singlets).

The Fayet--Iliopoulos parameters $\xi_P$  in (\ref{Dvev}) are determined by the 
quantum version of the classical expressions
(\ref{xis}) \cite{SYfstr,SYrvacua}.   
Defining
\beq
u_k= \left\bra {\rm Tr}\left(\frac12\, a + T^a\, a^a\right)^k\right\ket, \qquad k=1, ..., N\,,
\label{u}
\eeq
 we  obtain 
\cite{SYfstr}
\beq
\xi_P=-2\sqrt{2}\,\mu\,E_P\,,
\label{qxis}
\eeq
where $E_P$ ($P=1, ..., N$) are the diagonal elements of the $N\times N$ matrix
\beq
E=\frac1{N}\,\frac{\pt u_2}{\pt a}+T^{\tilde{a}}\,\frac{\pt u_2}{\pt a^{\tilde{a}}}\,,
\label{E}
\eeq
and $T^{\tilde{a}}$ are the Cartan generators of the SU$(N)$ gauge group ($\tilde{a}=1,...,(N-1)$).
The $E_P$ parameters  are expressible  via the roots of the Seiberg--Witten curve (see below). 

The Seiberg--Witten curve in  our theory takes the form \cite{APS}
\beq
y^2= \prod_{P=1}^{N} (x-\phi_P)^2 -
4\left(\frac{\Lambda_{{\mathcal N}=2}}{\sqrt{2}}\right)^{2N-N_f}\, \,\,\prod_{A=1}^{N_f} \left(x+\frac{m_A}{\sqrt{2}}\right).
\label{curve}
\eeq
Here $\phi_P$ are gauge invariant parameters on the Coulomb branch. Semiclassically,
\beq
{\rm diag}\left(\frac12\, a + T^a\, a^a\right) \approx 
\left[\phi_1,...,\phi_N\right].
\eeq
In the $r=N$ vacuum the curve (\ref{curve}) has $N$ double roots associated with condensation of $N$ quarks.
 It reduces to
\beq
y^2= \prod_{P=1}^{N} (x-e_P)^2,
\label{rNcurve}
\eeq
where quasiclassically (at large masses) $e_P$'s and $\phi_P$'s are given  by the
mass parameters, $\sqrt{2}e_P\approx \sqrt{2}\phi_P\approx -m_P$, $P=1, ... , N$.
In the  $r<N$ quark vacuum (i.e. the ($1, ... ,\, r$) vacuum) we have
\beq
\phi_P \approx -\frac{m_P}{\sqrt{2}},\qquad P=1, ... ,\, r\,, \qquad 
\phi_P \sim \Lambda_{{\mathcal N}=2},\qquad P=r+1, ... ,\, N\,
\label{classphi}
\eeq
in the large $m_A$ limit, see (\ref{avevr}).

To identify the $r<N$ vacuum in terms of the curve (\ref{curve}) it is necessary to find
such values of $\phi_P$ which ensure that the curve has $N-1$ double roots, and
$r$ parameters $\phi_P$ are determined by the quark masses in the semiclassical limit, see (\ref{classphi}).
$(N-1)$ double roots are associated with $r$ condensed quarks and $(N-r-1)$ condensed monopoles --
 altogether $N-1$ condensed states.
In contrast, in the $r=N$ vacuum the maximal possible number of condensed states (quarks) 
in the U$(N)$ theory is $N$. As was already mentioned, this difference is related to the 
the unbroken U(1)$^{\rm unbr}$ gauge group
in the $r<N$ vacua \cite{Cachazo2}. In the classically unbroken (after quark condensation) U$(N-r)$ gauge group,
$N-r-1$ monopoles condense at the quantum level breaking the non-Abelian SU$(N-r)$ subgroup. One U(1)
factor remains unbroken because monopoles do not interact with it.

Thus in the $r<N$ vacua the Seiberg--Witten curve factorizes \cite{CaInVa},
\beq
y^2
=\prod_{P=1}^{N-1} (x-e_P)^2\,(x-e_N^{+})(x-e_N^{-}).
\label{rcurve}
\eeq
The last two  roots (and $\phi_N$) are of order of 
$\Lambda_{{\mathcal N}=2}$. For the single-trace deformation superpotential (\ref{msuperpotbr}) 
corresponding to 
$\gamma=0$ (see (\ref{gamma})) their sum vanishes \cite{CaInVa},
\beq
e_N^{+} + e_N^{-}=0\,.
\label{DijVafa}
\eeq
This condition is equivalent to a physical condition
\beq
\xi_N= -2\sqrt{2}\mu\,E_N =0\,,
\label{xiN0}
\eeq
which ensures that the $N$-th quark is heavy and  develops no VEV \cite{SYrvacua}. 
The root $e_N^{+}$ determines the value of the gaugino condensate \cite{Cachazo2}, see (\ref{lambdacond})  
in Sec.~\ref{CSWsol}.

The parameters $E_P$
in the  $r=N$ vacuum   are given by double roots of the Seiberg--Witten curve \cite{SYfstr}, namely,
\beq
E_P=e_P, \qquad P=1, ..., N\,.
\label{ErN}
\eeq
This implies, in turn, that the dyon condensates at small $\xi$ in the $r=N$ vacuum are  
\beq
\xi_P=-2\sqrt{2}\,\mu\,e_P\,.
\label{xirN}
\eeq
As long as we keep $\xi_P$ small (i.e. in the domain (\ref{strcoup}))
the coupling constants of the
infrared-free dual theory (frozen at the scale of the dyon VEVs) are small;
the dual theory is at weak coupling.

At small $m_A-m_B\equiv\Delta m_{AB}$, in the domain  (\ref{strcoup}), the double roots of the Seiberg--Witten
 curve are  
\beq
\sqrt{2}e_I = -m_{I+N}, \qquad 
\sqrt{2}e_J = \Lambda_{{\mathcal N}=2}\,\exp{\left(\frac{2\pi i}{N-\tN}J\right)}
\label{roots}
\eeq
for $N-\tN>1$, where 
\beq
I=1, ... ,\tN\,\,\,\, {\rm  and} \,\,\,\, J=\tN+1, ... , N\,.
\label{d1}
\eeq
 In particular, the first $\tN$  roots are determined by the masses of the 
 last $\tN$ quarks --- a reflection of the fact that the 
non-Abelian sector of the dual theory is not asymptotically free and is at weak coupling
in the domain (\ref{strcoup}). 

In the $r<N$ vacua 
the relation between the parameters $E_P$ which determine 
the dyon condensates and the roots of the Seiberg--Witten curve changes. Namely, we have \cite{SYrvacua}
\beq
E_P=\sqrt{(e_P-e_N^{+})(e_P-e_N^{-})}, \qquad P=1,...,(N-1), \qquad E_N=0\,.
\label{Ee}
\eeq
In terms of the roots of the Seiberg-Witten curve this implies for the dyon VEVs 
\beq
\xi_P=-2\sqrt{2}\,\mu\,\sqrt{(e_P-e_N^{+})(e_P-e_N^{-})}, \quad P=1,...,(N-1), \quad \xi_N=0\,.
\label{xirN-1}
\eeq

In  much the same way as in the $r=N$ vacuum, the first   $\nu$ roots are determined by the masses of the 
 last $\nu$ quarks
at small $\Delta m_{AB}$, i.e. in the domain (\ref{strcoup}), 
\beq
\sqrt{2}e_I = -m_{I+r}, \qquad 
I=1, ... ,\nu    \,.
\label{nuroots}
\eeq
 This is because  the 
non-Abelian sector of the dual theory is infrared free and is at weak coupling
in the domain (\ref{strcoup}). 

\subsection{``Instead-of-confinement'' mechanism in the \\
 \boldmath{$r$} vacua}
\label{ICr}

The ``vacuum leap'' (\ref{jump}) ensures that in the $r$-vacua we have ``instead-of-confinement'' mechanism 
for quarks and gauge bosons  \cite{SYdual,SYrvacua} (assuming $\frac12 N_f< r\le N$).
Consider the mass choice (\ref{masssplit}), (\ref{masssplitr}).
Both, the gauge group and the global flavor SU($N_f$) group, are
broken in the vacuum. However, the color-flavor locked form of (\ref{Dvev})
shows that the  unbroken global group of the dual
theory is 
\beq
 {\rm SU}(r)_F\times  {\rm SU}(\nu)_{C+F}\times {\rm U}(1)\,.
\label{c+fdr}
\eeq
The SU$(\nu)_{C+F}$ factor in (\ref{c+fdr}) is a global unbroken color-flavor rotation, which involves the
last $\nu$ flavors, while the SU$(r)_F$ factor stands for the flavor rotation of the 
first $r$ dyons.

 In the equal mass limit, or given the mass choice (\ref{masssplit}), (\ref{masssplitr}),
the global unbroken symmetry (\ref{c+fdr}) of the dual theory at small
$\xi$ coincides with the global group (\ref{c+fr}) (or (\ref{c+f}) for $r=N$) in the
 the original theory at large $\xi$.  
However,  this global symmetry is realized in two different ways in the dual pair at hand.
The quarks and  gauge bosons of the original theory at large $\xi$
come in the $(1,1)$, $(r^2-1,1)$, $(\bar{r},\nu)$, and $(r,\bar{\nu})$
representations (see (\ref{onepr}), (\ref{twopr}) or (\ref{onep}), (\ref{twop})), while the dyons and U($\nu$) gauge 
bosons form 
\beq
(1,1), \qquad (1,\nu^2-1)
\label{adjdual}
\eeq
and 
\beq
 (r,\bar{\nu}), \qquad  (\bar{r},\nu)
\label{bifunddual}
\eeq 
representations of (\ref{c+fdr}). We see  that the
adjoint representations of the $(C+F)$
subgroup are different in two theories.
     
This means that quarks and gauge bosons
which form the  adjoint $(r^2-1)$ representation  
of SU($r$) at large $\xi$ and the dyons and dual gauge bosons which form the  adjoint $(\nu^2-1)$ representation  of SU($\nu$) at small $\xi$ are different states. What happens with quarks and gauge bosons
at small $\xi$?

Screened quarks 
and gauge bosons which exist in the $r$ vacuum in the large-$\xi$ domain
 decay into the monopole-antimonopole 
pairs on the CMS.\footnote{Strictly speaking,
such pairs  can be  formed by monopole-antidyons and
 dyon-antidyons as well, the dyons carrying root-like electric charges. 
 As was already explained above, we collectively refer to all such states  as 
``monopoles'' to avoid confusion with quark-like dyons which appear in Eq.~(\ref{Dvev}). The
latter dyons carry weight-like electric charges, see \cite{SYdual,SYrvacua} for details.}
This is in accordance with the results obtained 
in \ntwo SU(2) gauge theories \cite{SW1,SW2,BF} on the Coulomb branch at vanishing $\xi$;
 for the theory under consideration such a behavior was established in \cite{SYtorkink}.
The general rule is that the only states which exist at strong coupling inside CMS are those which can become massless on the Coulomb branch
\cite{SW1,SW2,BF}. For our theory these are light dyons shown in Eq.~(\ref{Dvev}),
gauge bosons of the dual gauge group and monopoles.

As shown in \cite{SYdual,SYrvacua}, at small nonvanishing $\xi$ the
monopoles and antimonopoles produced in the decay process of the adjoint $(r^2-1,1)$ states
 are confined. Therefore, the (screened) quarks or gauge bosons 
evolve into stringy mesons in the  strong coupling domain of small  $\xi$ 
-- the  monopole-antimonopole
pairs connected  by two strings, as shown in  Fig.~\ref{figmeson}. 

The distinction between the ``instead-of-confinement'' phase in 
the $r<N$ vacua and that  in the $r=N$ vacuum is that in the $r<N$ vacua
the
strings can be broken by $M_{PN}$-monopole-antimonopole pairs, $P=1,... , r$. Here $M_{PN}$ denote
the monopoles that are  junctions of the $P$-th   and $N$-th strings (the latter is spurious,
 see \cite{SYrvacua}
for details). As a
result, the dipole stringy states emitting unbroken U(1)$^{{\rm unbr}}$ magnetic gauge fields
are formed, see Fig.~\ref{figdipole}. Non-Abelian SU$(\nu)$ fluxes are confined in these stringy dipoles.

\section{\boldmath{$r$} Duality at large $\mu$}
\label{rduallargemu}
\setcounter{equation}{0}

In this section we discuss continuation of $r$ duality to the domain of large but finite $\mu$,
i.e.  \none SQCD.
We consider separately two cases: $r=N$   \cite{SYN1dual} and $r<N$  \cite{SYrvacua}.

\subsection{The \boldmath{$r=N$} vacuum}

From Eqs.~(\ref{Dvev}), (\ref{qxis}) and (\ref{roots}) we see that the VEVs of the non-Abelian
dyons $D^{lA}$ are determined by $\sqrt{\mu m}$ and are much smaller 
than the VEVs of the Abelian dyons $D^{J}$ at small $m$.
  The latter are of order of 
$\sqrt{\mu \Lambda_{{\mathcal N}=2}}$. 
This circumstance is most crucial for us. It  allows us to  increase $\mu$
and decouple the adjoint fields without 
ruining the weak coupling condition in the dual theory \cite{SYN1dual}.

Now we assume that 
\beq
|\mu| \gg |m_A |, \qquad A=1, ... , N_f\,.
\label{mularge}
\eeq
The Abelian dyon VEVs become large at large $\mu$. This makes  heavy
the U(1) gauge fields of the dual group 
(\ref{rdualgaugegroup}). Decoupling these gauge factors together with 
adjoint matter and the Abelian dyons themselves we get the low-energy theory with 
the gauge group
\beq
U(\tN)
\label{dualgglmurN}
\eeq
 and non-Abelian dyons $D^{lA}$ ($l=1,...,\tN$ and $A=1,...N_f$).
For the single-trace $\gamma=0$ perturbation (see (\ref{msuperpotbr})) the superpotential 
for $D^{lA}$ has the form \cite{SYN1dual}
\beq
{\mathcal W} = -\frac1{2\mu}\,
(\tilde{D}_A D^B)(\tilde{D}_B D^A)  
+m_A\,(\tilde{D}_A D^A)\, ,
\label{superpotd}
\eeq
where the color indices are contracted inside each parentheses.

Minimization of this superpotential leads to the dyon VEVs  shown in the first line
of Eq.~(\ref{Dvev}). Note, that 
 $\xi$'s which determine the non-Abelian dyon VEVs   are of  order $\mu m$, see (\ref{roots}).

Below the scale $\mu$ our theory becomes dual to \none SQCD with the scale
\beq
\tilde{\Lambda}_{r=N}^{N-2\tN} \,=\, \frac{\Lambda_{{\mathcal N}=2}^{N-\tN}}{\mu^{\tN}}\,.
\label{tildeL}
\eeq
The only condition we impose to keep this infrared-free theory at the weak coupling 
is 
\beq
|\sqrt{\mu m}| \ll \tilde{\Lambda}_{r=N}\,.
\label{wcdual}
\eeq
This means that at large $\mu$ we must keep the quark masses sufficiently small, which is always achievable.

We would like to stress that if all dyon VEVs were of order 
 of $\sqrt{\mu \Lambda_{{\mathcal N}=2}}$,
it would not be  possible to  decouple 
the adjoint matter keeping the dual theory at weak coupling. As soon as we 
increase $\mu$ beyond  the above  scale,  we will break the weak coupling 
condition in the dual theory.

\subsection{The \boldmath{$r<N$} vacua}

In order to keep our dual theory at weak coupling we need to constrain the  parameters 
$\xi$ (at least $\nu$ of them) from above. At large $\mu$ this creates a problem.  This problem was overcame in \cite{SYrvacua} as follows.
Equation (\ref{xirN-1}) shows that if we assume the mass differences 
to be very small and fine-tune the average value of $\nu$ double roots  (determined by masses,
which are almost equal) to be close to one of the roots $e_N^{\pm}$, we guarantee
  $\nu$ parameters $\xi$ to be  small.
Say, we tune the quark masses to ensure  
\beq
e_P \to e_N^{+}, \quad \Delta m_{KK'} \ll \Lambda_{{\mathcal N}=2}, \quad P=1,...,\nu, \quad K,K'= (r+1),...,N_f.
\label{ADlimit}
\eeq
Note, that it is possible to make all $\nu$ double roots close to $e_N^{+}$ because 
it is the quark masses rather 
than $\Lambda_{{\mathcal N}=2}$ that determine the ``non-Abelian''  roots of 
the Seiberg--Witten curve and VEVs of the non-Abelian dyons, see (\ref{nuroots}).

The above limit means moving towards the
Argyres--Douglas (AD) regime \cite{AD}. 
Indeed, on the Coulomb branch  the masses of $\nu$ monopoles $M_{PN}$
($P=1,...,\nu$) are determined by the differences $(e_P - e_N^{+})\to 0$; the corresponding 
$\beta$-cycles shrink. Thus, in addition to the  light dyons $D^{lA}$ and $D^J$ which are always present in our
$r$ vacuum we get extra light monopoles mutually non-local with dyons. If we were on the Coulomb branch 
(at $\xi_P=0$) this  would definitely mean moving into strong coupling. 

However, at small but nonvanishing $\xi$ we are 
{\em not} on the Coulomb branch. In fact, monopoles are confined. 
As  shown in \cite{SYrvacua},
this ensures  the theory to stay at weak coupling. Basically the reason is that
 $\nu$ monopoles $M_{PN}$, $P=1,... ,\nu$ in question form stringy dipole states shown
in Fig.~\ref{figdipole}. Although the $M_{PN}$ masses   themselves tend to zero in the limit 
(\ref{ADlimit}) the mass of the stringy dipole state formed by these monopole and antimonopole 
is determined by the string tension. It is of order of $\sqrt{\xi_P}$ and, therefore, is much larger.
This ensures the smallness of the  renormalized coupling constant, provided we keep $\xi$'s small enough.
The fact that the light matter VEVs tend to zero in the AD point was first recognized 
in \cite{GVY} in  the  Abelian case.

Now we can proceed in much  the same way as in the $r=N$ vacuum. We let $\mu$ grow passing 
to the  limit (\ref{ADlimit}).
The VEVs of the non-Abelian dyons $D^{lA}$ become much smaller than the  VEVs of 
the Abelian dyons $D^J$, see
(\ref{Dvev}), (\ref{xirN-1}) and (\ref{nuroots}). 
As a result,
 $(N-\nu-1)$ U(1) gauge fields of the dual gauge group (\ref{rdualgaugegroup}), together with $D^J$ dyons themselves, acquire large masses 
($\sim \sqrt{\mu \Lambda_{{\mathcal N}=2}}$) and decouple.
At large $\mu$,
\beq
|\mu| \gg |\sqrt{\xi}|
\label{muL}
\eeq
adjoint matter decouples too. 

 In order to keep the dual theory at weak coupling we 
go to the AD limit (\ref{ADlimit}) and require
\beq
|\sqrt{\xi_P}|\ll \tilde{\Lambda}_{r}, \qquad P=1,...,\nu\, ,
\label{weakcouprd}
\eeq
where 
\beq
\tilde{\Lambda}^{r-2\nu}_{r}= \frac{\Lambda_{{\mathcal N}=2}^{r-\nu }}{\mu^{\nu}}\,.
\label{tildeLr}
\eeq
We also assume that quark mass differences are very small, even smaller than $E_P$, namely,
\beq
\Delta m_{KK'} \ll E_P=\sqrt{(e_P^2-e_N^{2})},\quad P=1,...,\nu, \quad K,K'= (r+1),...,N_f\,.
\label{DeltamE}
\eeq
At low energies our dual theory has the gauge group
\beq
{\rm U}(\nu)\times {\rm U}(1)^{\rm unbr},
\label{dggrouplmu}
\eeq
while  light matter is represented by 
the non-Abelian dyons $D^{lA}$ ($l=1,...,\nu$ and  $A=1,...,N_f$). The superpotential is  \cite{SYrvacua}
\beqn
{\mathcal W} &=& \frac{\hat{E}}{\sqrt{2}\,\hat{m}\,\mu}\,
 (\tilde{D}_A D^B)(\tilde{D}_B D^A) +\left[(m_A-\hat{m})+\frac{(\sqrt{2}\,\hat{E})^2}{\hat{m}}\right]\,
(\tilde{D}_A D^A)
\nonumber\\[3mm]
&+& 
c\left[\frac1{2\mu}\,(\tilde{D}_A D^A)^2 + \sqrt{2}\nu\,\hat{E}\,(\tilde{D}_A D^A)\right],
\label{superpotrd}
\eeqn
where $c$ is a constant, $c\sim 1$. Here 
\beq
\hat{m}=\frac1{\nu}\,\, \sum_{P=1}^{\nu} m_{r+P}, \qquad 
\hat{E}=\frac1{\nu}\,\, \sum_{P=1}^{\nu} E_{P}\,= \frac1{\sqrt{2}}\,\sqrt{\hat{m}^2-\frac{4S}{\mu}},
\label{m}
\eeq
where $S$ is the gaugino condensate, see (\ref{lambdacond}).
The non-Abelian dyon VEVs  obtained from this superpotential are given by the first line in (\ref{Dvev}). They
are small, corresponding $\xi$'s are of order of $\mu\hat{E}$.

\subsection{Summary}

Systematizing the overall picture behind $r$ duality  (in \none, i.e. at large $\mu$, and 
under the condition $\frac23 N_f<r\le N$)   upon reducing $\xi$ the original theory
 undergoes  a crossover transition
at strong coupling. In the region (\ref{weakcouprd}) at small quark masses in the  $r=N$ vacuum (or close
to the AD points (\ref{ADlimit}) in the  $r<N$ vacua) the strongly coupled theory is described by a dual
weakly coupled infrared-free theory, U$(\tN)$ or U$(\nu)\times$U(1)$^{\rm unbr}$ SQCD, with   
$N_f$ light dyon flavors.  Condensation of light dyons
$D^{lA}$ in the dual theory leads to formation of non-Abelian strings and confinement of monopoles.
Quarks and gauge bosons
 of the original \none SQCD are in  the ``instead-of-confinement'' phase: they decay into the
 monopole-antimonopole pairs on CMS and 
form stringy mesons. In the $r<N$ vacua  in the AD-regime (\ref{ADlimit}), the $M_{PN}$ monopoles  ($P=1,... ,\nu$)
become very light and, therefore, strings are highly unstable. As a result,
the  stringy mesons shown in  Fig.~\ref{figmeson}
decay into stringy dipoles, see Fig~\ref{figdipole}. Stringy dipoles with nontrivial charges with respect to the
SU$(r)$ part of the global group (for example, adjoint) are stable.

\section{Generalized Seiberg's duality}
\label{seiberg}
\setcounter{equation}{0}

Now we would like to compare $r$ duality we  established with Seiberg's duality \cite{Sdual,IS}. 
Originally Seiberg's duality was formulated for \none SQCD   corresponding to the limit $\mu\to \infty$.
Therefore, in the original formulation Seiberg's duality referred to the monopole vacua with $r=0$.
Other vacua, with $r\ne 0$, have condensates of $r$ quark flavors  $\langle \tilde{q}q\rangle_{A} \sim \mu m_A$
and, therefore, become runaway vacua in the limit $\mu\to \infty$. 

At the same time, the $r$ duality \cite{SYrvacua} can be continued to large but finite $\mu$ in the $r$ vacua 
 ($\frac23 N_f < r\le N$), see Sec.~\ref{rduallargemu}.
In order to compare both dualities  with each other
we rely on a generalization of Seiberg's duality to the $r$ vacua \cite{CKM}.\footnote{It was suggested in  \cite{CKM} 
for SU$(N)$ gauge theories. We use here a similar formulation for U$(N)$ gauge theories. For a later development see
\cite{GivKut}.}

At large $\mu$ one can integrate out adjoint matter in superpotentials (\ref{superpot}), (\ref{msuperpotbr}).
For the single-trace deformation with $\gamma=0$ this gives the superpotential
\beq
-\frac1{2\mu}\,(\tilde{q}_A q^B)(\tilde{q}_B q^A) + m_A\,(\tilde{q}_A q^A),
\eeq
where color indices inside the brackets are contracted. 
This suggests that the Seiberg dual theory for our $\mu$-deformed U$(N)$
\ntwo SQCD at large but finite $\mu$  has the gauge group (\ref{dualgglmurN})
and $N_f$ flavors of Seiberg's ``dual quarks'' $h^{lA}$ ($l=1,...,\tN$ and $A=1,...,N_f$) and 
(being \none supersymmetric) possesses superpotential
\beq
{\cal W}_{S}= -\frac{\kappa^2}{2\mu}\,{\rm Tr}\,(M^2) + \kappa \,m_A\, M_A^A +\tilde{h}_{Al}h^{lB}\,M_B^A,
\label{Ssup}
\eeq
where $M_A^B$ is the Seiberg neutral mesonic $M$ field defined as
\beq
(\tilde{q}_A q^B)=\kappa\,M_A^B.
\label{M}
\eeq
Here $\kappa$ is a parameter of dimension of mass needed to formulate Seiberg's duality \cite{Sdual,IS}.

From the definition (\ref{M}) it is clear that 
in the $r$ vacuum the number of eigenvalues of the matrix $\tilde{q} q =\kappa M$ which scales as
$ \mu m$ at large $ \mu m$ is $r$. Moreover,
\beq
r\le N\,,
\label{rN}
\eeq
since classically the rank of the $(\tilde{q}_B q^A)$ matrix  cannot exceed $N$.

Now let us discuss the vacuum structure of the Seiberg dual theory (\ref{Ssup}). We do it separately in the
$r=N$ and $r<N$ vacua.

\subsection{The \boldmath{$r=N$} vacuum}

Let us minimize superpotential (\ref{Ssup}) to find the classical vacua of 
the generalized Seiberg dual theory. Assuming that $\langle M_A^B\rangle =\delta_A^B\,M_A$
we obtain the equations
\beqn
&&
-\frac{\kappa^2}{\mu}\,M_A + \kappa\,m_A +\tilde{h}_{Al}h^{lA}=0,
\nonumber\\
&&
M_A\,h^{lA}=\tilde{h}_{Al}\,M_A =0,
\label{veqs}
\eeqn
which should be valid for any $A$.

To solve these equations we note 
that the rank of the $\tilde{h}_{Ak}h^{kB}$ matrix  cannot exceed $\tN$. 
In particular, for $r=N$ vacuum we have the maximal number of condensed $h$-fields equal to $\tN$.
 In this case we can choose
the $(1,...,N)$ vacuum as follows
\beqn
&&
 M_A = \frac{\mu}{\kappa}\, m_A, \qquad (\tilde{h}h)_A=0, \qquad A=1,... , N\,,
\nonumber\\
&&
(\tilde{h}h)_A = -\kappa\,m_A, \qquad M_A=0, \qquad A=(N+1),... , N_f\,,
\label{r=Nvac}
\eeqn
where $(\tilde{h}h)_A$ are diagonal elements of the matrix $\tilde{h}_{Ak}h^{kB}$.
The number the of $r=N$ vacua is given in (\ref{numva}).
 It is equal to the  number of possibilities of  choosing $N$ nonvanishing elements $M_A$ out of $N_f$.
This is also the number of the  $r=N$ vacua in the original theory at small
$\mu$, i.e. close to the \ntwo limit.

Now we assume the fields $M_A^B$ to be heavy and integrate them out. This implies that $\kappa$ is 
large.\footnote{We will see that the parameter $\kappa$ does not enter the low-energy theory.}
Integrating out the $M$ fields in (\ref{Ssup}) we get
\beq
{\cal W}_{S}^{\rm LE} = \frac{\mu}{2\kappa^2}\, (\tilde{h}_{A}h^{B})(\tilde{h}_{B}h^{A}) + 
\frac{\mu}{\kappa}\,m_A\,(\tilde{h}_{A}h^{A})\,.
\label{SLEsup}
\eeq
The  change of variables
\beq
D^{lA}=\sqrt{-\frac{\mu}{\kappa}}\, h^{lA}, \qquad l=1,...,\tN, \qquad A=1,...,N_f
\label{change}
\eeq
brings this superpotential to the form 
\beq
{\cal W}_{S}^{\rm LE} = \frac1{2\mu}\,
(\tilde{D}_A D^B)(\tilde{D}_B D^A)  
-m_A\,(\tilde{D}_A D^A)\,.
\label{S=SY}
\eeq
We see that (up to a sign) this superpotential coincides   with the superpotential of our $r$ dual theory
(\ref{superpotd}). As was already mentioned, the
 dual gauge groups  also coincide for Seiberg's and $r$ dualities in the $r=N$ vacuum. Note, that kinetic terms are not known in the Seiberg's dual theory, thus, normalization of the $h$ fields is  not   fixed.

This leads us to the conclusion
  that in the  $r=N$ vacua both dual theories are identical. In Appendix A we show that this 
  coincidence remains valid for more generic (non-single-trace) deformations of \ntwo SQCD.  Of course, upon  identification (\ref{change}) the $h^{lA}$  VEVs  (\ref{r=Nvac}) coincide with the VEVs of the $D^{lA}$
  dyons  in (\ref{Dvev}) in the $r=N$ vacuum, see (\ref{xirN}) and (\ref{roots}).

The identification (\ref{change}) reveals the physical nature of Seiberg's ``dual quarks''.
They are not monopoles as naive duality suggests. Instead,
 they are quark-like dyons appearing in the $r$-dual theory below crossover. Their condensation leads to confinement of monopoles and ``instead-of-confinement'' phase for the quarks and gauge bosons of the original theory.


\subsection{The \boldmath{$r<N$} vacua}

As we will show in Sec.~\ref{GK}  there are no classical supersymmetric vacua in the Seiberg dual theory
with superpotential (\ref{Ssup}) for $r$ vacua in the range $\frac23 N_f<r<N$. 
However, one can look for quantum vacua.
Following \cite{IS,IS2}, we assume that the $M_A^B$ fields develop VEVs making ``dual quarks'' heavy and 
then integrate 
$h^{lA}$ out.
The gluino condensation in the U$(\tN)$ gauge theory with no matter induces the superpotential
\beq
{\mathcal W}_{S}^{\rm eff} = -\frac{\kappa^2}{2\mu}\,{\rm Tr}\,(M^2) + \kappa \,m_A\, M_A^A +\tN\,\tilde{\Lambda}_{S}^{\frac{2\tN -N}{\tN}} \left({\rm det}\,M\right)^{\frac{1}{\tN}},
\label{quS}
\eeq
where $\tilde{\Lambda}_{S}$ is the scale of Seiberg's dual theory defined via \cite{Sdual,IS}
\beq
\tilde{\Lambda}_{S}^{N-2\tN}\Lambda^{2N-\tN}=(-1)^{\tN}\,\kappa^{N_f},
\label{LambdaS}
\eeq 
while $\Lambda$ is the scale of the original \none theory. It is related to $\Lambda_{{\mathcal N}=2}$ as 
follows:
\beq
\Lambda^{2N-\tN} = \mu^{N}\,\Lambda_{{\mathcal N}=2}^{N-\tN}
\label{Lambda}
\eeq
Substituting the definition of the $M$ fields (\ref{M}) in (\ref{quS}) we arrive at
\beq
{\mathcal W}_{S}^{\rm eff} = -\frac{1}{2\mu}\,{\rm Tr}\,(\tilde{q}q)^2 + m_A\, {\rm Tr}\,(\tilde{q}q )
+(N-N_f)\,\frac{\Lambda^{\frac{3N-N_f}{N-N_f}}}{\left({\rm det}\,\tilde{q}q\right)^{\frac1{N-N_f}}}
\,.
\label{ads}
\eeq
The last quantum term  is nothing other than the continuation of the
Afleck--Dine--Seiberg (ADS) superpotential 
\cite{ADS} to $N_f>N$. As was explained in the beginning of this section,
 the first term is obtained by integrating out adjoint fields in the original theory. 
 Note, that in much the same way as in the $r=N$ vacuum the dependence on
the $\kappa$  parameter disappeared. 

Assuming, as before, that the matrix $(\tilde{q}_A q^B)$ is diagonal we present
 the vacuum equation in the form
\beq
\frac1{\mu}\,(\tilde{q}q)_A = m_A -\frac1{\Lambda^{\frac{2N-\tN}{\tN}}}\,
\frac{\prod_B \left[(\tilde{q}q)_B\right]^{\frac1{\tN}}}{(\tilde{q}q)_A}\,.
\label{veq}
\eeq
The superpotential (\ref{ads}) is exact and we can use it in any domain in the parameter space. In particular,
for large masses, $m_A\gg \Lambda_{{\mathcal N}=2}$, the solution of Eq.~(\ref{veq}) for
the  $(1,...,r)$ vacuum
is
\beqn
&&
(\tilde{q}q)_A \approx  \mu\, m, \qquad A=1,...,r
\nonumber\\[2mm]
&&
(\tilde{q}q)_A \approx  \mu\, \Lambda_{{\mathcal N}=2}^{\frac{N-\tN}{N-r}}\;
m^{\frac{\tN-r}{N-r}}\;e^{\frac{2\pi k}{N-r}\,i}, \qquad A=(r+1),...,N_f, 
\nonumber\\[2mm]
&&
 k=1,... , (N-r)\,,
\label{qvevS}
\eeqn
where we assume the equal mass limit for simplicity. We have $r$ large classical VEVs and $(N_f-r)$ small 
``quantum'' VEVs.

The linear dependence of $(\tilde{q}q)$ on $\mu$ is exact and is fixed by U(1) symmetries \cite{GVY}
after condensates are expressed in terms of $\Lambda_{{\mathcal N}=2}$. The presence of 
$(N-r)$ solutions ensures that the total number of the $r<N$ vacua in our theory is
\beq
{\cal N}_{r<N}=\sum_{r=0}^{N-1} \,(N-r)\,C_{N_f}^{r}= \sum_{r=0}^{N-1}\, (N-r)\,\frac{N_f!}{r!(N_f-r)!},
\label{nurvacS}
\eeq
where the upper limit for $r$ is implemented by the condition (\ref{rN}).
This number coincides with the result (\ref{nurvac}) obtained in the \ntwo limit and, therefore, matches  number 
of the $r<N$ vacua in our $r$-dual theory.\footnote{Our large-$m$ counting (\ref{nurvacS}) also agrees with the final result in \cite{CKM}, see also Sec.~\ref{GK}.}

\subsection{Generalized Seiberg's duality and exact chiral rings}
\label{CSWsol}

In this section 
we would like to make sure that generalized Seiberg's duality gives 
correct values of the chiral condensates in the $r<N$ vacua.
To this end we compare quark condensates determined by Eq.~(\ref{ads}) with 
the exact results obtained in \cite{Cachazo2}.
This section  overlaps  with what is already known, and we include it here mostly for the sake of completeness.
For example, a somewhat similar analysis for SU$(N)$ gauge theory can be found in \cite{Pietro}.

All chiral condensates in our theory can be encoded in the following functions \cite{Cachazo2}
\beqn
&&
T(x)=\left\langle {\rm Tr}\,\frac1{x-\Phi}\right\rangle,
\nonumber\\[2mm]
&&
R(x)=\frac1{32\pi^2}\,\left\langle {\rm Tr}\,\frac{W_{\alpha}W^{\alpha}}{x-\Phi}\right\rangle,
\nonumber\\[2mm]
&&
M(x)_A^B=\left\langle \tilde{q}_A\,\frac1{x-\Phi}\,q^B\right\rangle,
\label{rings}
\eeqn
where $W_{\alpha}$ is the gauge field strength superfield.
For the quadratic single-trace deformation (\ref{msuperpotbr}) (``one-cut'' model) the function
$R(x)$ has the form
\beq
R(x)=\frac12\left(W_{\rm br}^{'}(x) -\sqrt{W_{\rm br}^{'}(x)+f(x)}\right)
=\mu\,\left(x-\sqrt{x^2-e_N^2}\right),
\label{R(x)}
\eeq
where the undoubled root of the Seiberg--Witten curve $e_N = e_N^{+}$ (see (\ref{rcurve})) is related to the gaugino condensate,
\beq
e_N^2=\frac{2S}{\mu}, \qquad S=\frac1{32\pi^2}\,\left\langle {\rm Tr}\,W_{\alpha}W^{\alpha}\right\rangle.
\label{lambdacond}
\eeq
The solutions for the chiral rings were obtained in \cite{Cachazo2} in the $r<N$ vacua. In the
 $r=N$ vacuum the gaugino condensate vanishes, all roots of the 
 Seiberg--Witten curve are doubled, and there are no cuts in the $x$-plane.
As we already mentioned, all $r<N$ vacua belong to a single ``branch'' with a single U(1) gauge factor unbroken, while 
in the $r=N$ vacuum the gauge group is fully Higgsed.

From the solution for the function $M_A^B(x)$ in \cite{Cachazo2} one can obtain the values of 
the quark VEVs
in terms of the gaugino condensate $S$.
In the $r$ vacuum $(1,...,r)$ (when the function $M_A^B(x)$ has $r$ poles on the first sheet) we have
\beqn
&&
(\tilde{q}q)_A =  \frac{\mu}{2}\, \left(m_A+\sqrt{m_A^2-\frac{4S}{\mu}}\right), \qquad A=1,...,r
\nonumber\\[2mm]
&&
(\tilde{q}q)_A =  \frac{\mu}{2}\, \left(m_A-\sqrt{m_A^2-\frac{4S}{\mu}}\right), \quad A=(r+1),...,N_f, 
\label{qVEVcsw}
\eeqn

\vspace{2mm}

Now, to find the gaugino condensate $S$ we use the glueball superpotential calculated in \cite{Cachazo2} from
a matrix model \cite{DijVafa}. For our theory with the quadratic single-trace deformation (\ref{msuperpotbr})
it has the form \cite{Ookouchi} 
\beqn
&&
{\cal W}_{\rm glueball}= S\,\left[N+\log{\frac{\mu^N\,\Lambda_{{\mathcal N}=2}^{N-\tN}\,
\prod_{A} m_A}{S^N}}\right]
\nonumber\\[3mm]
&&
-\sum_{A=r+1}^{N_f} S\left[-\log{\left(\frac12 +\frac12\sqrt{1-\frac{4S}{\mu m_A^2}}\right)}
+\frac{\mu m_A^2}{4S}\left(\sqrt{1-\frac{4S}{\mu m_A^2}}-1\right) +\frac12\right]
\nonumber\\[3mm]
&&
-\sum_{A=1}^{r}S\left[-\log{\left(\frac12 -\frac12\sqrt{1-\frac{4S}{\mu m_A^2}}\right)}
 \right.
\nonumber\\[3mm]
&&
\left.
+\frac{\mu m_A^2}{4S}\left(-\sqrt{1-\frac{4S}{\mu m_A^2}}-1\right)+\frac12\right].
\label{CSWsup}
\eeqn
Minimization of this superpotential gives the equation for $S$ from which we obtain
\beq
S^N= \mu^N\,\Lambda_{{\mathcal N}=2}^{N-\tN}\,\left(\frac{m}{2}-\frac12\sqrt{m^2-\frac{4S}{\mu}}\right)^r
\,\left(\frac{m}{2}+\frac12\sqrt{m^2-\frac{4S}{\mu}}\right)^{N_f-r},
\label{Seqn}
\eeq
where we assume the equal-mass limit for simplicity.

Now let us derive equations for the quark VEVs using 
the Cachazo--Seiberg--Witten expressions (\ref{qVEVcsw}) and 
equation (\ref{Seqn}). To this end we first express the right-hand side of (\ref{Seqn}) in terms of 
the quark VEVs
using (\ref{qVEVcsw}). Solving this equation for $S$ we get
\beq
S=\frac{\left({\rm det}\,\tilde{q}q\right)^{\frac1{\tN}}}
{\mu^{\frac{N}{\tN}}\,\Lambda_{{\mathcal N}=2}^{\frac{N-\tN}{\tN}}}\,\,.
\label{623}
\eeq
Substituting $S$ from (\ref{623})  in the right-hand side
 of (\ref{qVEVcsw}) we derive the following equation for the quark VEVs:
\beq
\frac1{\mu}\,(\tilde{q}q)_A = m -\frac1{\mu^{\frac{N}{\tN}}\,\Lambda_{{\mathcal N}=2}^{\frac{N-\tN}{\tN}}}\,
\frac{\left({\rm det}\,\tilde{q}q\right)^{\frac1{\tN}}}{(\tilde{q}q)_A}.
\label{qeqCSW}
\eeq
These equations coincides with  those in (\ref{veq}) (for equal quark masses). We see that  
the Cachazo--Seiberg--Witten
exact solution \cite{Cachazo2} produces the same equations for 
the quark condensates as the  continuation of 
the ADS superpotential to $N_f>N$ in Eq.~(\ref{ads}). This justifies the latter superpotential.

\section{Classical and quantum \boldmath{$r$} vacua in Seiberg's dual theory}
\label{GK}
\setcounter{equation}{0}

As was mentioned in Sec.~\ref{intro},  generalized Seiberg's duality suggested in \cite{CKM} was later studied 
 in \cite{GivKut} in the \none theories with the
 SU$(N)$ gauge group.
The numbers of classical and quantum vacua corresponding to the superpotentials (\ref{Ssup}) and (\ref{ads})
were analyzed. In particular, a certain number of classical vacua was detected.  

In this section we show that 
that there are no classical vacua in the Seiberg's dual theory in the range $\frac23 N_f< r<N$ we explore
 in this paper. For smaller values of $r$, namely for $r<\tN$, the generalized Seiberg
  superpotential (\ref{Ssup}) does have classical vacua.

First, we briefly  review the analysis carried out in  \cite{GivKut}. The solution for (\ref{veqs}) was written
as (cf. (\ref{r=Nvac}))
\beqn
&&
 (\tilde{q}q)_A=\kappa M_A = \mu\, m_A, \qquad (\tilde{h}h)_A=0, \qquad A=1,... , p\,,
\nonumber\\[2mm]
&&
(\tilde{h}h)_A = -\kappa\,m_A, \qquad M_A=0,  \qquad A=(p+1),... , N_f\,,
\label{GKclvac}
\eeqn
where now $p$ should obey the constraint $p>N$, since the rank of the matrix $(\tilde{h}h)$ cannot exceed $\tN$,
and we do not consider the $r=N$ vacua in this section.

This solution can describe low-energy physics if the infrared-free Seiberg dual theory is at
weak coupling. To ensure this we assume the quark masses to be small, 
\beq
m_A\ll \Lambda_{{\mathcal N}=2}\,.
\label{smallmass}
\eeq
As we will see below, in the case at hand $p$  does not coincide with $r$, the latter  parameter being 
defined at large masses.
Therefore, the condition (\ref{rN}) does not apply for $p$.
The number of the above vacuum solutions is
\beq
{\cal N}^{S}_{\rm class}=\sum_{p=N+1}^{N_f} \,(p-N)\,C_{N_f}^{p}\,,
\label{nuclvacGK}
\eeq
where $(p-N)$ is the rank of the gauge group unbroken by the $h$-condensation, and 
we modify the results of \cite{GivKut} to include the combinatorial factor $C_{N_f}^{r}$, 
see  \cite{CKM}.\footnote{At this point we keep the
quark masses slightly different so that all vacua are isolated and can be counted. 
The calculation of \cite{GivKut} refers to the equal-mass limit and, in fact, corresponds 
to counting the number of the Higgs branches which are continuously
degenerate in the equal-mass limit (vacuum moduli).} The number of these classical vacua is less than the total 
number of 
the $r<N$ vacua
(\ref{nurvacS}). The missing vacua are in fact quantum vacua which are not seen in Seiberg's superpotential
(\ref{Ssup}) even at small $m_A$. They can be recovered from (\ref{ads}), however \cite{CKM,GivKut}.

The $(\tilde{q}q)$ matrix in Eq.~(\ref{veq}) has two different eigenvalues (in the limit of equal quark masses),
namely
\beq
(\tilde{q}q)^A_B= {\rm diag} (z,...,z,y,...,y),
\label{zy}
\eeq
where (at small $m$) $z$ appears $p$ times  while $y$ appears $(N_f-p)$ times,
and, in addition, 
\beq
z+y=\mu m\, .
\label{z+y}
\eeq
 From (\ref{veq}) we can write the following equation \cite{GivKut}:
\beq
z^{p-\tN}=\left(\mu \Lambda_{{\mathcal N}=2}\right)^{N-\tN}\,y^{p-N}
\label{xy}
\eeq
which, in combination with   (\ref{z+y}), allows us to determine both $z$ and $y$.

\vspace{1mm}

Following \cite{GivKut} we note that for $p\ge N$ Eq.~(\ref{xy}) is a polynomial of degree $(p-\tN)$
with respect to $z$ and, therefore, has $(p-\tN)$ solutions for $z$. For $\frac12 N_f\le p< N$ this equation has
$(N-\tN)$ solutions. Summing up all solutions together we get the number of vacua in the form
\beqn
&&
{\cal N}_{r<N}=\sum_{p=N+1}^{N_f} \,(p-\tN)\,C_{N_f}^{p} +  (N-\tN)\,\left(\sum_{p=\frac{N_f}{2}+1}^{N} \,C_{N_f}^{p} + \frac12 C_{N_f}^{\frac12 N_f}\right) 
\nonumber\\
&&
= \sum_{p=N}^{N_f} \,(p-N)\,C_{N_f}^{p}
+\frac12 (N-\tN)\,\sum_{p=0}^{N_f} \,C_{N_f}^{p}\,.
\label{nuvacGK}
\eeqn
The last sum here reduces to $(N-\tN)\,2^{N_f-1}$; and then Eq.~(\ref{nuvacGK}) can be rewritten as \cite{CKM}
\beq
{\cal N}_{r<N}=\sum_{p=N}^{N_f} \,(p-N)\,C_{N_f}^{p}+\sum_{p=0}^{N_f} \,(N-p)\,C_{N_f}^{p}=
\sum_{r=0}^{N} \,(N-r)\,C_{N_f}^{r}\,.
\label{vacnum}
\eeq

This calculation refers to the small-mass limit. The first term corresponds to the  number 
of the classical vacua (\ref{nuclvacGK}), while the second one counts the {\em missing quantum} vacua. 
The total number of vacua obviously
coincides with Eq.~(\ref{nurvacS}) obtained at large $m$.

\vspace{1mm}

Now, we can solve Eqs.~(\ref{z+y}) and (\ref{xy}) at small $m$. In addition to ``large'' solutions with
$z\approx -y \sim \Lambda_{{\mathcal N}=2}$, we also get ``small'' solutions
\beqn
&&
(\tilde{q}q)_A =z \approx  \mu\, m, \qquad A=1,...,p
\nonumber\\[2mm]
&&
(\tilde{q}q)_A =y \approx  \mu\, 
\frac{m^{\frac{p-\tN}{p-N}}}{\Lambda_{{\mathcal N}=2}^{\frac{N-\tN}{p-N}}}\;e^{\frac{2\pi k}{p-N}\,i}, \qquad A=(p+1),...,N_f, 
\nonumber\\[2mm]
&&
 k=1,... , (p-N)\,.
\label{smallqvev}
\eeqn
These solutions should be compared with the classical solutions (\ref{GKclvac}). We see that the $m$
dependence
of $(\tilde{q}q)$ matches; thus, these solutions corerespond to 
the classical vacua of Seiberg's
dual theory. In order to have $y$ much smaller than $z$ in the small-mass limit we impose the
condition
$p>N$. (This is to be contrasted with  the condition $r<N$ in (\ref{qvevS}) for large $m$). Given the multiplicity of these
solutions equal to $(p-N)$ we see that the number of these vacua precisely matches the number of 
the classical
Seiberg vacua (\ref{nuclvacGK}).

The behavior of $(\tilde{q}q)$ in (\ref{smallqvev}) ensures that the gaugino condensate is very small
in these vacua, see (\ref{qVEVcsw}). Namely,
\beq
S\approx  \mu\, 
\frac{m^{\frac{2p-N_f}{p-N}}}{\Lambda_{{\mathcal N}=2}^{\frac{N-\tN}{p-N}}}\;e^{\frac{2\pi k}{p-N}\,i}\,, \qquad 
 k=1,... , (p-N)\,.
\label{gauginosmallm}
\eeq

\vspace{1mm}

What is the relation between $r$ and $p$? 

\vspace{1mm}

At large $m_A$ we start  from an $r$ vacuum,  with $r$ quarks (classically) condensed, hence
$r\le N$. On the other hand, $p$ is defined as 
the number of ``plus'' signs in Eq.~(\ref{qVEVcsw}) for $$(\tilde{q}q)_A=z\,.$$
(Then $(N_f-p)$ is the number of ``minus'' signs). In fact, $p$ depends on $m_A$. At large $m_A$ we have
$p(\infty)=r$. As  we reduce $m_A$ certain poles can pass through the cut from the first sheet to the second or
vice versa \cite{Cachazo2}. When it happens $p(m_A)$ reduces by one unit or increases by one unit. 

In Eqs.~(\ref{smallqvev}) and (\ref{gauginosmallm})
$p$ is $p(m_A)$ in the small mass limit, $p=p(0)$. Clearly $p$ can   differ  from $r$, and 
the condition (\ref{rN})
does not apply for $p$. In fact, $(p-r)$
is the net number of poles which pass through the cut from the second sheet to the first one as we reduce 
the quark masses
from infinity to zero. 

At large $m$ we start in the  $r$ vacuum, with $r<N$, and 
the quark condensate given by (\ref{qvevS}). This solution corresponds to 
\beq
S\approx  \mu\,
\Lambda_{{\mathcal N}=2}^{\frac{N-\tN}{N-r}}\;
m^{\frac{N_f-2r}{N-r}}\;e^{\frac{2\pi k}{N-r}\,i}, \qquad 
 k=1,... , (N-r)\,.
\label{gauginolargem}
\eeq
This behavior can be seen in  Eq.~(\ref{Seqn}) 
 as follows. We expand the square roots in $S/\mu\,m^2$ in the right-hand side of (\ref{Seqn}).
The second factor tends to a constant while the first factor gives $S^r$, which reproduces the behavior in 
(\ref{gauginolargem}).

Now, to determine the relation between $r$ and $p$ in the vacua which are described by Seiberg's duality 
we must find the solution of Eq.~(\ref{Seqn}) which approaches (\ref{gauginolargem}) at large $m$ and has 
the behavior (\ref{gauginosmallm}) at $m\to 0$. 

There is  only one possibility for this to happen.  As $m$ reduces all poles should pass thorough 
the cut, so that the signs of the 
square roots in (\ref{Seqn}) change. In other words, as we reduce $m$ from large to small 
values, all $r$ poles from the first sheet pass to the second one and, simultaneously, all $(N_f-r)$ poles from the second sheet pass
 to the first one. Then at small $m$
the first factor in the right-hand side of Eq.~(\ref{Seqn}) tends to a constant, while the second one gives
$S^{N_f-r}$. This gives
 the behavior (\ref{gauginosmallm}) where
\beq
p=N_f-r\,.
\label{pn}
\eeq

We stress that there are other solutions to (\ref{Seqn}) which have different behavior at small $m$ ($S\sim \mu\,
\Lambda_{{\mathcal N}=2}^2$). We are interested in the behavior (\ref{gauginosmallm}) with $p>N$ because these
solutions correspond to the
vacua seen classically in the Seiberg dual theory. Other vacua are ``quantum''
vacua (see (\ref{vacnum})) which remain classically invisible.

For Seiberg's  classical vacua we need $p> N$. This translates into the constraint
\beq
r<\tN.
\label{smallr}
\eeq
In this paper we study $r$-duality in the range $\frac23 N_f< r\le N$; thus,
 the above vacua are beyond the range of our  analysis. This means 
 that the $r$ vacua   described by our $r$-duality should be interpreted as ``missing quantum'' vacua
from the standpoint of Seiberg's duality. 

\section{\boldmath{$r$} Duality versus Seiberg's duality for \\the \boldmath{$\frac23 N_f<r<N$} vacua}
\label{CSW}
\setcounter{equation}{0}

For  $\frac23 N_f<r<N$ vacua our $r$-dual theory    does not agree with the generalized Seiberg
dual theory.
First, we have the  U$(\nu)$ gauge group, while the Seiberg
dual  has the gauge group U$(\tN)$. Light matter sectors  and effective superpotentials are also
different in two theories: the $D^{lA} $ dyons  ($l=1,...,\nu$) with superpotential (\ref{superpotrd}) versus ``dual quarks'' $h^{lA} $ ($l=1,...,\tN$) plus $M$ fields with superpotential (\ref{Ssup}). 

Both dual theories are well justified
and verified. On the one hand, the $r$-dual theory is derived from the \ntwo limit
by increasing $\mu$ and keeping the theory at weak coupling at all intermediate stages. On the other hand, as was 
checked in Sec.~\ref{CSWsol}, the  generalized Seiberg dual theory (more exactly, the generalized ADS superpotential
(\ref{ads})) matches  the exact solution of \cite{Cachazo2}. What is going on?

Our interpretation is as follows.
In the $r$-vacua (in the range $\frac{2}{3}N_f < r < N$) the  generalized Seiberg dual theory is at strong 
coupling and, therefore,  
cannot describe low-energy physics in its entirety. 
However, it does describe the chiral sector in the sense of the
Veneziano--Yankielowicz effective superpotential \cite{Ven}.  Namely, 
condensates from the chiral ring are correctly reproduced. 
 The spectrum of excitations is not.

As an example consider superpotentials (\ref{ads}) or (\ref{CSWsup}). Although these 
superpotentials correctly reproduce 
the chiral condensates, taken at their face value  
they  do not describe low energy-spectrum. Namely, it is clear that neither
the quark mesonic field $(\tilde{q}q)$ nor 
the glueball field $S$ are light degrees of freedom at strong coupling.

We believe that the superpotential (\ref{Ssup}) is of the same kind 
in the window $\frac23 N_f<r<N$. This assertion is supported by the fact that  
supersymmetric  vacua are not seen at the classical level 
in the superpotential (\ref{Ssup}) for $\frac23 N_f<r<N$.  
In order to find supersymmetric vacua we have to integrate out the $h$ fields 
and search for  solutions in the effective quantum superpotential (\ref{ads}). 
This suggests that the ``dual quarks'' $h$ are  not the low-energy degrees of freedom 
and, in fact, Seiberg's dual theory (\ref{Ssup}) is strongly coupled at  small $\xi$'s 
in this window.

Instead, the $r$-dual theory {\em is} the low-energy description at small $\xi$, where the original \none SQCD is at strong coupling. As long as we keep the parameters $\xi$ small (see (\ref{weakcouprd})) 
the $r$-dual infrared-free theory is at weak coupling and under control.  Condensation of 
the quark-like dyons $D^{lA}$ in this theory leads
to  confinement of monopoles and ``instead-of-confinement'' phase for 
the quarks and gauge bosons.

As was shown in Sec.~\ref{GK}, in the range $r < \tN$ the generalized Seiberg's dual theory has supersymmetric 
classical vacua and, being infrared-free, is at weak coupling  (the same applies to
the $r=N$ vacuum where it matches $r$-duality). Therefore, it does describe low-energy physics in the $r$ vacua
with $r < \tN$. The schematic picture of both dual descriptions versus $r$ is shown in 
Fig.~\ref{figdual}.

\vspace{2mm}

{\sl A very important problem for future studies is  extrapolating $r$ duality to $r\le \frac{2}{3}N_f$ and comparing it 
in this range with Seiberg's duality. Another problem  is understanding $r$ duality in the framework of strings/branes, in the spirit
it had been done with the Seiberg duality.}

\section*{Acknowledgments}

We thank Zohar Komargodski for a very helpful discussion.
Useful correspondence with Amit Giveon and David Kutasov
is gratefully acknowledged.

 The work of MS was supported in part by DOE
grant DE-FG02-94ER40823. 
The work of AY was  supported
by  FTPI, University of Minnesota
and by Russian State Grant for
Scientific Schools RSGSS-65751.2010.2.

\section*{Appendix:   
Non-single-trace deformations in the
\boldmath{$r=N$} vacuum}

\renewcommand{\theequation}{A.\arabic{equation}}
\setcounter{equation}{0}
 
 \renewcommand{\thesubsection}{A.\arabic{subsection}}
\setcounter{subsection}{0}

In this Appendix we show the matching of the effective superpotentials for $r$-dual and generalized Seiberg's dual theories for generic deformation (\ref{msuperpotbrg}) with $\gamma\neq 0$. For this case the superpotential
of $r$-dual theory obtained in \cite{SYN1dual} has the form
\beqn
{\mathcal W} &=& -\frac1{2\mu}\,
\left[ (\tilde{D}_A D^B)(\tilde{D}_B D^A) - \frac{\alpha_{D}}{\tN} (\tilde{D}_A D^A)^2 \right]\, 
\nonumber\\[3mm]
&+& 
\left[m_A-\frac{\gamma\,(1+\frac{\tN}{N})}{1+\gamma\,\frac{\tN}{N}}\,m\right]\,(\tilde{D}_A D^A),
\label{superpotdgamma}
\eeqn
where 
\beq
\alpha_{D}=\frac{\gamma\,\frac{\tN}{N}}{ 1+\gamma\,\frac{\tN}{N}}, \qquad 
m=\frac1{N_f}\,\, \sum_{A=1}^{N_f} m_A \,
\label{alphaD}
\eeq
and $\gamma$ is given by (\ref{gamma}).

On the other hand the generalized  Seiberg's  superpotential for $\gamma\neq 0$ is
\beq
{\cal W}_{S}= -\frac{\kappa^2}{2\mu}\,{\rm Tr}\,(M^2) +\frac{\kappa^2}{2N\mu}\,\alpha\,{(\rm Tr}\,M)^2)
+ \kappa \,m_A\, M_A^A +\tilde{h}_{Al}h^{lB}\,M_B^A,
\label{Ssupgamma}
\eeq
where 
\beq
\alpha= 1-\sqrt{\frac{N}{2}}\,\frac{\mu}{\mu_0}=-\frac{\gamma}{1-\gamma}.
\label{alpha}
\eeq
Upon integrating out the $M$ fields we get
\beqn
&&
{\cal W}_{S}^{\rm LE} = \frac{\mu}{2\kappa^2}\, \left[(\tilde{h}_{A}h^{B})(\tilde{h}_{B}h^{A})
- \frac{\alpha_{D}}{\tN} (\tilde{h}_A h^A)^2\right]
\nonumber\\
&&
+ \frac{\mu}{\kappa}\,\left[m_A  - \frac{\gamma\,(1+\frac{\tN}{N})}{1+\gamma\,\frac{\tN}{N}}\,m\right]\,(\tilde{h}_{A}h^{A}).
\label{SLEsupgamma}
\eeqn

We see again that upon change of variables (\ref{change}) two superpotentials (\ref{SLEsupgamma}) and 
(\ref{superpotdgamma}) coincide (up to a sign).

\end{document}